\documentclass{article}
\usepackage{graphicx}
\usepackage{jcappub}

\title{Testing an Inflation Model with Nonminimal Derivative Coupling in the Light of PLANCK 2015 Data}

\author{Kourosh Nozari,}

\author{Narges Rashidi}

\affiliation{Department of Physics, Faculty of Basic Sciences,\\
University of Mazandaran,\\
P. O. Box 47416-95447, Babolsar, IRAN}

 \emailAdd{knozari@umz.ac.ir}

\emailAdd{n.rashidi@umz.ac.ir}

\abstract{We study the dynamics of a generalized inflationary model
in which both the scalar field and its derivatives are coupled to
the gravity. We consider a general form of the nonminimal derivative
coupling in order to have a complete treatment of the model. By
expanding the action up to the second order in perturbation, we
study the spectrum of the primordial modes of the perturbations.
Also, by expanding the action up to the third order and considering
the three point correlation functions, the amplitude of the
non-Gaussianity of the primordial perturbations is studied both in
equilateral and orthogonal configurations. Finally, by adopting some
sort of potentials, we compare the model at hand with the Planck
2015 released observational data and obtain some constraints on the
model's parameters space. As an important result, we show that the
nonminimal couplings help to make models of chaotic inflation, that
would otherwise be in tension with Planck data, in better agreement
with the data. This model is consistent with observation at weak
coupling limit.}


\begin{document}

\maketitle


\section{Introduction}

The existence of an inflationary stage in the history of the early
universe was proposed to solve some problems of the standard big
bang cosmology, such as the flatness, horizon and relics problems. A
successful inflationary paradigm can also provide a causal mechanism
for generating the initial density perturbations needed to seed the
formation of structures in the universe
~\citep{Gut81,Lin82,Alb82,Lin90,Lid00a,Lid97,Rio02,Lyt09}. A simple
single field inflationary scenario (in which the early Universe was
dominated by the potential energy of the scalar field dubbed
inflaton) predicts the dominant mode of the primordial density
fluctuations to be highly adiabatic, scale invariant and
Gaussian~\citep{Mal03}. But, recent observational data have detected
a level of scale dependence in the primordial
perturbations~\citep{Ade15a,Ade15b}. On the other hand, although
there is no direct signal for primordial non-Gaussianity in
observation, Planck team has obtained some tight limits on
primordial non-Gaussianity~\citep{Ade15c}. Some inflationary models
also, predict a level of non-Gaussianity in the perturbation's
mode~\citep{Mal03,Bar04,Che10,Fel11a,Fel11b,Noz12,Noz13a,Noz13b,Noz13c}.
This is really an important and interesting point. Because, a large
amount of information on the cosmological dynamics deriving the
initial inflationary expansion of the Universe can be found in the
primordial scalar non-Gaussianities~\citep{Bab04a,Che08}. In this
regard, the extended inflationary models, which can show the
non-Gaussianity and scale dependence of the primordial perturbation,
are more favorable.

One successful class of the extended inflationary models is the one
with a nonminimal coupling between the scalar field and the
curvature scalar $R$. This coupling usually is shown by the
lagrangian term as $f(\phi)R$. There are some motivations behind
including an explicit nonminimal coupling in the action. It is
necessary for the renormalization term arising in quantum field
theory in curved space~\citep{Bir82}. In considering the quantum
corrections to the scalar field theory, this term would arises at
the quantum level ~\citep{Far96,Far00}. With nonminimal coupling
term, it is possible to have gravity as a spontaneous
symmetry-breaking effect~\citep{Zee79,Acc85}. NMC term allows to
have an oscillating Universe~\citep{Ran84}. Also, with this term, it
is possible to solve the graceful exit problem of the old inflation
model by slowing the false-vacuum expansion~\citep{La89a,La89b}. So,
it seems that the presence of nonminimal coupling between the scalar
field and Ricci scalar is an useful extension of the theory and in
this regard many authors have studied such
models~\citep{Noz12,Fut89,Sal89,Fak90,Mak91,Hwa99,Tsu00b,
Pal11,Noz07b,Noz10}.

It has been shown that the presence of NMC makes the inflation model
more viable. For instance, the author of Ref.~\citep{Lee14} has
considered an inflation model with nonminimal coupling and
non-canonical kinetic terms leading to an chaotic inflation model.
It is shown that this chaotic inflation model is favored by the
BICEP2 observation. But this model is in tension with recent
observational data released by Planck2015. The authors of
Ref.~\citep{Gla15} have considered an inflation model with a
nonminimal coupling between the scalar field and gravity as
$(\alpha\phi^{2}+\beta\phi^{4})R$. They compared their model with
Planck2015 data in Einstein frame to explore the viability of the
model. It is shown that this model just for $N=65$ and with some
maximum values of $n_{s}$ is consistent with Planck2015 data. In
Ref.~\citep{Bou15} a nonminimally coupled inflaionary model with
quadratic potential in Einstein frame has been considered. The
authors of this paper have shown that their model for some values of
the NMC parameter is consistent with observational data. For other
literature dealing with these issues we refer the reader to
Refs.~\citep{Mos14,Pal15,Dal14,Lin14a,Lin15,Kal15,Kal14a,Kal14b,Lin14b,Kal14c}.

Another extension of the inflation theory arises from considering a
nonminimal coupling between the derivatives of the scalar field and
the curvature, which leads to the interesting cosmological
behaviors~\citep{Ame93}. This coupling is usually given by the
lagrangian term as
$\left(R_{\mu\nu}-\frac{1}{2}g_{\mu\nu}R\right)\nabla^{\mu} \phi
\nabla^{\nu}\phi$. Some authors have shown that during inflation,
with a nonminimal coupling between the scalar field and gravity, the
unitary bound of the theory violates~\citep{Bur09,Bar09,Bur10}.
However, it has been shown that the model with a nonminimal
couplings between the kinetic term of the inflaton (derivatives of
the scalar field) and the Einstein tensor preserves the unitary
bound during inflation~\citep{Ger10}. Also, the presence of
nonminimal derivative coupling is a powerful tool to increase the
friction of an inflaton rolling down its own
potential~\citep{Ger10}. Some authors have considered the model with
this coupling term and have studied the early time accelerating
expansion of the universe as well as the late time
dynamics~\citep{Tsu12,Sad2013,Sar10}. As an important extension of
the nonminimal models, in addition to the standard coupling between
the scalar field and curvature, it is possible also to have coupling
between the derivatives of the scalar field and gravitational sector
which makes the inflationary model more powerful to be a successful
scenario.

In this work, we consider a generalized inflationary model in which
both the scalar field and its derivatives are coupled to the
curvature. We also, consider an extension of the nonminimal
derivative term in the lagrangian as
${\cal{F}}(\phi)G_{\mu\nu}\nabla^{\mu}\nabla^{\nu}(\phi)$. In this
term, ${\cal{F}}$ is a general function of $\phi$ which for
${\cal{F}}\sim\frac{1}{2}\phi$, leads to the simple cases introduced
earlier in literature. In section 2 we present some main equations
of this generalized inflationary model. In section 3, we use the ADM
metric and study the linear perturbations of the model. By expanding
the action up to the second order in perturbation, and considering
the 2-point correlation functions, we obtain the amplitude of the
scalar perturbation and its spectral index. Also, by considering the
tensor part of the perturbed metric, we obtain the tensor
perturbation and its spectral index as well. In section 4, by
expanding the action up to the cubic order in perturbation, we
explore the non-linear perturbation in this model. By considering
the 3-point correlation functions, we study the non-Gaussian modes
of the primordial perturbations. In this section we obtain the
amplitude of the non-Gaussianity in the equilateral and orthogonal
configuration and in $k_{1}=k_{2}=k_{3}$ limit. Then, in section 5,
we test our generalized inflationary model in confrontation with the
recently released observational data. To this end, we consider two
types of nonminimal coupling parameter ( $\xi\phi^{2}$ and
$\xi\phi^{4}$), several types of nonminimal derivative coupling
function and also, several types of potential. By these functions we
study the behavior of the perturbative parameters in the background
of the Planck 2015 observational data. Our main goal is to see
whether this non-minimal model has the potential to make models of
chaotic inflation in better agreement with Planck 2015 data or not.

In this regard, the model with non-minimal coupling between the
scalar field and its derivative to gravity is a subset of Horndeski
theory which satisfies the necessary conditions for being a good
theoretical inflationary model. Also, this model is observationally
viable, since this model with several inflationary potential is
consistent with Planck2015 observational data. As we can see from
Planck team's released paper, a simple inflationary model, with
potential corresponding to $\phi^{2}$, $\phi^{3}$ and
$\phi^{\frac{2}{3}}$ is not consistent with observation. In
Ref.~\citep{Kal14c} it has been shown that a nonminimal model (a
model with nonminimal coupling between the scalar field and gravity)
is viable with potentials they have considered. However, their model
has been studied in Einstein frame. We will show that adding a
nonminimal coupling between the gravity and derivatives of the
scalar field leads to a viable inflationary model even in Jordan
frame.

The main properties of the cosmological perturbations are described
by the tensor-to-scalar ratio and the scalar spectral index. In this
regard, several observational teams are trying to find some
constraints on these perturbative parameters. WMAP team has found
the constraints $r<0.13$ and $n_{s}=0.9636\pm 0.0084$ from the
combined WMAP9+eCMB+BAO+H$_{0}$ data~\citep{Hin13}. The constraints
obtained by the joint Planck 2013+WMAP9+BAO data are as $r<0.12$ and
$n_{s}=0.9643\pm 0.0059$~\citep{Ade13}. Then, in 2014, BICEP team
found a surprising result. The joint Planck 2013+WMAP+BICEP2+BAO
data has set the constraints $r=0.2096^{+0.0443}_{-0.0608}$ and
$n_{s}=0.9653\pm 0.0129$ on the perturbative
parameters~\citep{Ade14,Wu14}. This large value of the
tensor-to-scalar ratio, which shows the large amplitude of the
gravitational wave modes generated during inflation, was a really
challenging result. But Planck collaboration has performed a new
study on the polarized dust emission in our galaxy and has shown
that the part of the sky observed by BICEP2 is contaminated
significantly by galactic-dust emission. So, the polarized dust
emission might contribute a significant part of the B-mode detected
by BICEP2. The Planck collaboration released the constraints
$r<0.099$ and $0.9652\pm 0.0047$ from Planck TT, TE, EE+lowP +WP
data (note that Planck TT, TE, EE+lowP refer to the combination of
the likelihood at $l>30$ using TT, TE, and EE spectra and the low
$l$ multipole temperature polarization
likelihood)~\citep{Ade15a,Ade15b,Ade15c}. In this regard, to compare
our model with observational data, we study the behavior of the
tensor-to-scalar ratio versus the scalar spectral index in the
background of the Planck TT, TE, EE+lowP data and deduce some
constraints on the parameters space of the model. We also analyze
the non-Gaussianity feature of the model by studying the behavior of
the orthogonal configuration versus the equilateral configuration in
the background of the observational data.

\section{The Setup}
The four-dimensional action for a cosmological model where a scalar
field is nonminimally coupled to the Ricci scalar, in the presence
of a nonminimal derivative coupling between the scalar field and
gravity, is given by the following expression
\begin{eqnarray}
\label{eq1} S=\int
d^{4}x\sqrt{-g}\Bigg[\frac{1}{2\kappa^{2}}R+f(\phi)R-\frac{1}{2}\partial_{\mu}\phi
\partial^{\mu}\phi-V(\phi)
+{\cal{F}}(\phi)G_{\mu\nu}\nabla^{\mu}\nabla^{\nu}(\phi)\Bigg]\,,\hspace{0.6cm}
\end{eqnarray}
where, $R$ is the Ricci scalar, $\phi$ is a scalar field (inflaton)
with the potential $V(\phi)$ and $f(\phi)$ and ${\cal{F}}(\phi)$ are
general functions of the scalar field. Also, in the nonminimal
derivative term of the action \eqref{eq1},
$G_{\mu\nu}=R_{\mu\nu}-\frac{1}{2}g_{\mu\nu}R$ is the Einstein
tensor. Note that, in some papers there is a coefficient
$\frac{\kappa^{*}}{2}$ in front of the nonminimal derivative term,
where the constant parameter $\kappa^{*}$ has dimension of
length-squared. In this paper, we absorb such a coefficient in
function ${\cal{F}}(\phi)$. By assuming the FRW metric, variation of
the action \eqref{eq1} with respect to the metric leads to the
Friedmann equation of the model as
\begin{equation}
\label{eq2}
H^{2}\left(1+2\kappa^{2}f\right)=\frac{\kappa^{2}}{3}\left(\dot{\phi}^{2}(\frac{1}{2}-9H^{2}{\cal{F}}')-6Hf'\dot{\phi}+V(\phi)\right),
\end{equation}
where, a dot refers to a time derivative of the parameter and a
prime denotes a derivative with respect to the scalar field. By
varying the action \eqref{eq1} with respect to the scalar field we
get the following equation of motion
\begin{eqnarray}
\label{eq3}
\ddot{\phi}\Big(-1+6{\cal{F}}'H^{2}\Big)+\Big(12{\cal{F}}'H\dot{H}+18{\cal{F}}'H^{3}-3H\Big)\dot{\phi}
+3{\cal{F}}''H^{2}\dot{\phi}^{2}+6f'R-V'=0\,.\hspace{0.6cm}
\end{eqnarray}

Now, by defining the slow-roll parameters as
$\epsilon\equiv-\frac{\dot{H}}{H^{2}}$ and
$\eta=-\frac{1}{H}\frac{\ddot{H}}{\dot{H}}$, we find these
parameters in our setup as follows
\begin{equation}
\label{eq4} \epsilon=\frac{A}{1+2\kappa^{2}f+\kappa
{\cal{F}}'\dot{\phi}^{2}},
\end{equation}
\begin{eqnarray}
\label{eq5}
\eta=2\epsilon-\frac{\dot{A}}{H\epsilon(1+2\kappa^{2}f+\kappa
{\cal{F}}'\dot{\phi}^{2})}
+\frac{A}{H\epsilon}\frac{2\kappa^{2}f'\dot{\phi}+\kappa^{2}{\cal{F}}''\dot{\phi}^{2}+\kappa^{2}{\cal{F}}'\ddot{\phi}}
{(1+2\kappa^{2}f+\kappa {\cal{F}}'\dot{\phi}^{2})^{2}},
\end{eqnarray}
where parameter $A$ is defined as
\begin{eqnarray}
\label{eq6}
A\equiv\frac{\kappa^{2}\dot{\phi}^{2}}{2H^{2}}-\frac{\kappa^{2}f'\dot{\phi}}{H}-3\kappa^{2}{\cal{F}}'\dot{\phi}^{2}
+\frac{\kappa^{2}f''\dot{\phi}^{2}}{H^{2}}+\frac{\kappa^{2}{\cal{F}}''\dot{\phi}^{3}}{H}
+\frac{\ddot{\phi}}{H\dot{\phi}}\Big(\frac{\kappa^{2}f'\dot{\phi}}{H}+\frac{\kappa^{2}{\cal{F}}'\dot{\phi}^{2}}{H}\Big)\,.\hspace{0.5cm}
\end{eqnarray}
During the inflationary era, the evolution of the Hubble parameter
is so slow, so that in this era the conditions $\epsilon \ll 1$ and
$\eta \ll 1$ are satisfied. As one of these two slow-varying
parameters reaches unity, the inflation phase terminates.

The number of e-folds during inflation which is defined as
\begin{equation}
\label{eq7} N=\int_{t_{hc}}^{t_{f}} H dt\,,
\end{equation}
in our setup and within the slow-roll limit ($\ddot{\phi}\ll
|3H\dot{\phi}|$ and $\dot{\phi}^{2}\ll V(\phi)$) takes the following
form
\begin{equation}
\label{eq8} N\simeq \int_{\phi_{hc}}^{\phi_{f}} \frac{3H^{2}
\Big(4{\cal{F}}'\dot{H}+6{\cal{F}}'H^{2}-1\Big)}{V'-6f'R} d\phi\,,
\end{equation}
where $\phi_{hc}$ shows the value of the inflaton at the horizon
crossing of the universe scale and $\phi_{f}$ denotes the value of
the field when the universe exits the inflationary phase. In the
next section, we study the linear perturbation of the model which is
the key test of any inflationary model. In this regard, we calculate
the spectrum of perturbations produced due to quantum fluctuations
of the fields about their homogeneous background values.

\section{Linear Perturbation}
In this section, we study the linear perturbation arising from the
quantum behavior of both the scalar field $\phi$ and the space-time
metric, $g_{\mu\nu}$, around the homogeneous background solution. To
this end, we should expand the action \eqref{eq1} up to the second
order in small fluctuations. In this regard, it is convenient to
work in ADM metric formalism~\citep{Arn60}
\begin{equation}
\label{eq9}
ds^{2}=-N^{2}dt^{2}+h_{ij}\big(dx^{i}+N^{i}dt\big)\big(dx^{j}+N^{j}dt\big),
\end{equation}
where $N$ is the lapse function and $N^{i}$ is the shift vector. To
obtain a general perturbed form of the metric \eqref{eq9}, we should
expand the shift and laps functions as $N^{i}\equiv
B^{i}=\delta^{ij}\partial_{j}B+v^{i}$ and $N=1+\Phi$, respectively.
$\Phi$ and $B$ are 3-scalar and $v^{i}$ is a vector satisfying the
condition $v^{i}_{,i}=0$~\citep{Muk92,Bau09}. Note that it is
sufficient to compute $N$ or $N^{i}$ up to the first order. This is
because the second order perturbation is multiplied by a factor
which is vanishing using the first order solution. Since the third
order term is multiplied by a constraint equation at the zeroth
order obeying the equations of motion, the contribution of this term
also vanishes~\citep{Mal03,Fel11a,Koy10}. $h_{ij}$ should be written
as
$h_{ij}=a^{2}\left[(1-2{\Psi})\delta_{ij}+2{\cal{T}}_{ij}\right]$,
where ${\Psi}$ is the spatial curvature perturbation and
${\cal{T}}_{ij}$ is a spatial shear 3-tensor which is symmetric and
traceless. So, the perturbed metric \eqref{eq9} takes the following
form
\begin{eqnarray}
\label{eq10} ds^{2}= -(1+2\Phi)dt^{2}+2a(t)B_{i}\,dt\,dx^{i}
+a^{2}(t)\left[(1-2{\Psi})\delta_{ij}+2{\cal{T}}_{ij}\right].
\end{eqnarray}
In order to study the scalar perturbation of the theory, we work
within the uniform-field gauge with $\delta\phi=0$ and also the
gauge ${\cal{T}}_{ij}=0$. So, by considering the scalar part of the
perturbations at the linear level, we rewrite the perturbed metric
\eqref{eq10} as~\citep{Muk92,Bau09,Bar80}
\begin{equation}
\label{eq11}
ds^{2}=-(1+2\Phi)dt^{2}+2a(t)B_{,i}\,dt\,dx^{i}+a^{2}(t)(1-2{\Psi})\delta_{ij}.
\end{equation}
Replacing the metric \eqref{eq11} in action \eqref{eq1} and
expanding the action up to the second order in perturbations gives
\begin{eqnarray}
\label{eq12} S_{2}=\int dt\,d^{3}x\,
a^{3}\Bigg[-3(\kappa^{-2}+f+\dot{\phi}^{2}{\cal{F}}')\dot{\Psi}^{2}-
\frac{2(\kappa^{-2}+f+\dot{\phi}^{2}{\cal{F}}')}{a^{2}}\Phi
\partial^{2}{\Psi}+\frac{1}{a^{2}}\Big(2(\kappa^{-2}+f+\dot{\phi}^{2}{\cal{F}}')\dot{\Psi}
\nonumber\\
-(2\kappa^{-2}H(1+\kappa^{2}f)+2\dot{\phi}f'+6\dot{\phi}^{2}{\cal{F}}')\Phi\Big)\partial^{2}B
+3\Big(2\kappa^{-2}H(1+\kappa^{2}f)+2\dot{\phi}f'+6\dot{\phi}^{2}{\cal{F}}'\Big)\Phi\dot{\Psi}
\nonumber\\
-\Big(3\kappa^{-2}H^{2}(1+\kappa^{2}f)-\frac{1}{2}\dot{\phi}^{2}+6H\dot{\phi}f'+18H^{2}\dot{\phi}^{2}{\cal{F}}'\Big)\Phi^{2}
+\frac{\kappa^{-2}+f-\dot{\phi}^{2}{\cal{F}}'}{a^{2}}(\partial{\Psi})^{2}\Bigg]\,.\hspace{1cm}
\end{eqnarray}

In this gauge, the equations of motion for $\Phi$ and $B$, obtained
by varying the action \eqref{eq12}, lead to the following constraint
equations
\begin{equation}
\label{eq13}
\Phi=2\frac{\kappa^{-2}+f+\dot{\phi}^{2}{\cal{F}}'}{2\kappa^{-2}H(1+\kappa^{2}f)+2\dot{\phi}f'+6\dot{\phi}^{2}{\cal{F}}'}\dot{\Psi},
\end{equation}
\begin{eqnarray}
\label{eq14}
\frac{\partial^{2}B}{a^{2}}=3\dot{\Psi}-\frac{2(\kappa^{-2}+f+\dot{\phi}^{2}{\cal{F}}')}{a^{2}(2\kappa^{-2}H(1+\kappa^{2}f)+2\dot{\phi}f'
+6\dot{\phi}^{2}{\cal{F}}')}
\partial^{2}{\Psi}\nonumber\\-\frac{2(9\kappa^{-2}H^{2}(1+\kappa^{2}f)
-\frac{3}{2}\dot{\phi}^{2}+18H\dot{\phi}f'+54H^{2}\dot{\phi}^{2}{\cal{F}}')}{3(2\kappa^{-2}H(1+\kappa^{2}f)+2\dot{\phi}f'
+6\dot{\phi}^{2}{\cal{F}}')}\Phi. \nonumber\\
\end{eqnarray}
By substituting equation \eqref{eq13} in action \eqref{eq12} and
integrating by parts, we reach
\begin{equation}
\label{eq15} S_{2}=\int
dt\,d^{3}x\,a^{3}{\cal{W}}\left[\dot{\Psi}^{2}-\frac{c_{s}^{2}}{a^{2}}(\partial
{\Psi})^{2}\right],
\end{equation}
where the parameters ${\cal{W}}$ and $c_{s}^{2}$ (which is known as
the sound velocity) are defined as
\begin{eqnarray}
\label{eq16}
{\cal{W}}=-4\frac{\left(\kappa^{-2}+f+\dot{\phi}^{2}{\cal{F}}'\right)^{2}\left(9\kappa^{-2}H^{2}(1+\kappa^{2}f)
-\frac{3}{2}\dot{\phi}^{2}+18H\dot{\phi}f'+54H^{2}\dot{\phi}^{2}{\cal{F}}'\right)}{\left(2\kappa^{-2}H(1+\kappa^{2}f)
+2\dot{\phi}f'+6\dot{\phi}^{2}{\cal{F}}'\right)^{2}}\nonumber\\
+3\left(\kappa^{-2}+f+\dot{\phi}^{2}{\cal{F}}'\right),
\end{eqnarray}

\begin{eqnarray}
\label{eq17}
c_{s}^{2}=\hspace{14.5cm}\nonumber\\3\Bigg[2\Big(2\kappa^{-2}H(1+\kappa^{2}f)
+2\dot{\phi}f'+6\dot{\phi}^{2}{\cal{F}}'\Big)\Big(\kappa^{-2}+f+\dot{\phi}^{2}{\cal{F}}'\Big)^{2}H
-\Big(2\kappa^{-2}H(1+\kappa^{2}f)
+2\dot{\phi}f'+6\dot{\phi}^{2}{\cal{F}}'\Big)^{2}
\nonumber\\
\Big(\kappa^{-2}+f-\dot{\phi}^{2}{\cal{F}}'\Big)+4\Big(2\kappa^{-2}H(1+\kappa^{2}f)
+2\dot{\phi}f'+6\dot{\phi}^{2}{\cal{F}}'\Big)\Big(\kappa^{-2}+f+\dot{\phi}^{2}{\cal{F}}'\Big)
\frac{d\Big(\kappa^{-2}+f+\dot{\phi}^{2}{\cal{F}}'\Big)}{dt}
\nonumber\\
-2\Big(\kappa^{-2}+f+\dot{\phi}^{2}{\cal{F}}'\Big)^{2}\,
\frac{d(2\kappa^{-2}H(1+\kappa^{2}f)
+2\dot{\phi}f'+6\dot{\phi}^{2}{\cal{F}}')}{dt}\Bigg] \Bigg[
\Bigg(9\Big(2\kappa^{-2}H(1+\kappa^{2}f)
+2\dot{\phi}f'\nonumber\\
+6\dot{\phi}^{2}{\cal{F}}'\Big)^{2}-4\Big(\kappa^{-2}+f+\dot{\phi}^{2}{\cal{F}}'\Big)\Big(9\kappa^{-2}H^{2}(1+\kappa^{2}f)
-\frac{3}{2}\dot{\phi}^{2}+18H\dot{\phi}f'+54H^{2}\dot{\phi}^{2}{\cal{F}}'\Big)
\Bigg)\Bigg]^{-1},
\end{eqnarray}

respectively. To see more details of driving this type of equations,
one can refer to Refs.~\citep{Fel11a,Fel11b,Che08,See05}.

Now we calculate the quantum perturbations of ${\Psi}$. To this end,
by variation of the action \eqref{eq15}, we find the equation of
motion of ${\Psi}$ as
\begin{equation}
\label{eq18}
\ddot{\Psi}+\left(3H+\frac{\dot{\cal{W}}}{\cal{W}}\right)\dot{\Psi}+c_{s}^{2}\,\frac{k^{2}}{a^{2}}\,{\Psi}=0.
\end{equation}
The solution of the above equation, up to the lowest order of the
slow-roll approximation, is given by the following expression
\begin{equation}
\label{eq19}
{\Psi}=\frac{iHe^{-ic_{s}^{2}k\tau}}{2c_{s}^{\frac{3}{2}}\sqrt{k^{3}{\cal{W}}}}\left(1+ic_{s}^{2}k\tau\right).
\end{equation}
To study the power spectrum of the curvature perturbation, we should
compute the two point correlation function in our setup. We find the
two-point correlation function, sometimes after the horizon
crossing, by obtaining the vacuum expectation value of ${\Psi}$ at
$\tau=0$ (corresponding to the end of inflation)
\begin{equation}
\label{eq20} \langle
0|{\Psi}(0,\textbf{k}_{1}){\Psi}(0,\textbf{k}_{2})|0\rangle
=(2\pi)^{3}\delta^{3}(\textbf{k}_{1}+\textbf{k}_{2})\frac{2\pi^{2}}{k^{3}}{\cal{A}}_{s},
\end{equation}
where ${\cal{A}}_{s}$ is called the power spectrum and is given by
\begin{equation}
\label{eq21}
{\cal{A}}_{s}=\frac{H^{2}}{8\pi^{2}{\cal{W}}c_{s}^{3}}\,.
\end{equation}
The scalar spectral index of the perturbations at the Hubble
crossing is defined as
\begin{equation}
\label{eq22} n_{s}-1=\frac{d \ln {\cal{A}}_{s}}{d \ln
k}\Bigg|_{c_{s}k=aH},
\end{equation}
which in our setup takes the following form
\begin{eqnarray}
\label{eq23} n_{s}-1=
-2\epsilon-\frac{5\kappa^{2}f'\dot{\phi}}{2H(1+\kappa^{2}f)}-\frac{1}{H}\frac{d
\ln
(\epsilon+\frac{5\kappa^{2}f'\dot{\phi}}{4H(1+\kappa^{2}f)})}{dt}
-\frac{1}{H}\frac{d \ln c_{s}}{dt}\,.\hspace{1cm}
\end{eqnarray}
Any deviation of $n_{s}$ from the unity shows that the evolution of
perturbations is scale dependent.

Other important parameters in an inflationary model are amplitude of
the tensor perturbation and its spectral index. To study the tensor
perturbation we use the tensor part of the perturbed metric
\eqref{eq10}. We can write ${\cal{T}}_{ij}$, in terms of the two
polarization tensors, as follows
\begin{equation}
\label{eq24}
{\cal{T}}_{ij}={\cal{T}}_{+}e_{ij}^{+}+{\cal{T}}_{\times}e_{ij}^{\times},
\end{equation}
where $e_{ij}^{(+,\times)}$ are symmetric, traceless and transverse.
The normalization condition imposes the constraints
\begin{equation}
\label{eq25}
e_{ij}^{(+,\times)}(\textbf{k})\,e_{ij}^{(+,\times)}(-\textbf{k})^{*}=2,
\end{equation}
\begin{equation}
\label{eq26}
e_{ij}^{(+)}(\textbf{k})\,e_{ij}^{(\times)}(-\textbf{k})^{*}=0,
\end{equation}
and the reality condition gives
\begin{equation}
\label{eq27}
e_{ij}^{(+,\times)}(-\textbf{k})=\left(e_{ij}^{(+,\times)}(\textbf{k})\right)^{*}.
\end{equation}
Now, we can write the second order action for the tensor mode
(gravitational waves) as follows
\begin{eqnarray}
\label{eq28} S_{T}=\int dt\, d^{3}x\, a^{3}
{\cal{W}}_{T}\left[\dot{\cal{T}}_{+}^{2}-\frac{c_{\cal{T}}^{2}}{a^{2}}(\partial
{\cal{T}}_{+})^{2}+\dot{\cal{T}}_{\times}^{2}-\frac{c_{\cal{T}}^{2}}{a^{2}}(\partial
{\cal{T}}_{\times})^{2}\right],\nonumber\\
\end{eqnarray}
where ${\cal{W}}_{T}$ and $c_{T}^{2}$ are defined with the following
expressions respectively
\begin{equation}
\label{eq29} {\cal{W}}_{T}=\frac{1}{4\kappa^{2}}(1+\kappa
f)(1+\frac{\kappa^{2}{\cal{F}}'\dot{\phi}^{2}}{1+\kappa f}),
\end{equation}
\begin{equation}
\label{eq30}
c_{T}^{2}=\frac{\kappa^{-2}+f-\dot{\phi}^{2}{\cal{F}}'}{\kappa^{-2}+f+\dot{\phi}^{2}{\cal{F}}'}\,.
\end{equation}
By following the strategy as applied for the scalar mode case, we
can find the amplitude of the tensor perturbations as follows
\begin{equation}
\label{eq31}
{\cal{A}}_{T}=\frac{H^{2}}{2\pi^{2}{\cal{W}}_{\cal{T}}c_{T}^{3}},
\end{equation}
which, by using the definition of the tensor spectral index
\begin{equation}
\label{eq32} n_{T}=\frac{d \ln {\cal{A}}_{T}}{d \ln k},
\end{equation}
leads to the following expression for $n_{T}$
\begin{equation}
\label{eq33}
n_{T}=-2\epsilon-\frac{\kappa^{2}f'\dot{\phi}}{2H(1+\kappa^{2}f)}.
\end{equation}
The ratio between the amplitudes of the tensor and scalar
perturbations (tensor-to-scalar ratio) is another important
inflationary parameter which in our setup is given by
\begin{equation}
\label{eq34}
r=\frac{{\cal{A}}_{T}}{{\cal{A}}_{T}}=16c_{s}\left(\epsilon+\frac{5\kappa^{2}f'\dot{\phi}}{4H(1+\kappa^{2}f)}\right).
\end{equation}

\section{Nonlinear Perturbations and Non-Gaussianity}

Non-Gaussianity of the primordial density perturbations is another
important aspect of an inflationary model. To explore the
non-Gaussianity of the density perturbations, one has to study the
nonlinear perturbation theory. The two-point correlation function of
the scalar perturbations gives no information about the
non-Gaussianity of the model. The non-Gaussian property shows itself
up in the three-point correlation function. This is because for a
Gaussian perturbation, all odd $n$-point correlators vanish and the
higher even $n$-point correlation functions can be expressed in
terms of sum of products of the two-point functions. To calculate
the three-point correlation function, we should expand the action
\eqref{eq1} up to the cubic order in the small fluctuations around
the homogeneous background solution. The obtained cubic terms, lead
to a change both in the ground state of the quantum field and
non-linearities in the evolution. By expanding the action
\eqref{eq1} up to the third order in perturbation, we obtain a
complicated expression which can be seen in the Appendix~\ref{A}. We
use equation \eqref{eq13} to eliminate the perturbation parameter
$\Phi$ in the expanded action. Then by introducing the new parameter
${\cal{X}}$ as
\begin{eqnarray}
\label{eq35}
B=\frac{2(\kappa^{-2}+f+\dot{\phi}^{2}{\cal{F}}'){\Psi}}{2\kappa^{-2}H(1+\kappa^{2}f)+2\dot{\phi}f'+6\dot{\phi}^{2}{\cal{F}}'}
+\frac{a^{2}{\cal{X}}}{\kappa^{-2}+f+\dot{\phi}^{2}{\cal{F}}'}\,,\hspace{1cm}
\end{eqnarray}
and
\begin{equation}
\label{eq36}
\partial^{2}{\cal{X}}={\cal{W}}\dot{\Psi}\,,
\end{equation}
we can rewrite the third order action, up to the leading order, as
follows

\begin{eqnarray}
\label{eq37} S_{3}=\int dt\, d^{3}x\,\Bigg\{
\Bigg[-\frac{3a^{3}}{\kappa^{2}}\,
\Bigg(\frac{1+\kappa^{2}f}{c_{s}^{2}}\Bigg)\Bigg(\frac{1}{c_{s}^{2}}-1\Bigg)
\left(\epsilon+\frac{5\kappa^{2}f'\dot{\phi}}{4H(1+\kappa^{2}f)}\right)
\Bigg]{\Psi}\dot{\Psi} +\Bigg[\frac{a}{\kappa^{2}}\,
\Bigg(1+\kappa^{2}f\Bigg) \nonumber\\
\Bigg(\frac{1}{c_{s}^{2}}-1\Bigg)\left(\epsilon+\frac{5\kappa^{2}f'\dot{\phi}}{4H(1+\kappa^{2}f)}\right)
\Bigg]{\Psi}\,(\partial{\Psi})^{2}+\Bigg[\frac{a^{3}}{\kappa^{2}}\,\Bigg(\frac{1+\kappa^{2}f}{c_{s}^{2}\,H}\Bigg)
\Bigg(\frac{1}{c_{s}^{2}}-1\Bigg)\left(\epsilon+\frac{5\kappa^{2}f'\dot{\phi}}{4H(1+\kappa^{2}f)}\right)\Bigg]
\dot{\Psi}^{3}\nonumber\\-\Bigg[a^{3}\,\frac{2}{c_{s}^{2}}\left(\epsilon+\frac{5\kappa^{2}f'\dot{\phi}}{4H(1+\kappa^{2}f)}\right)\dot{\Psi}
(\partial_{i}{\Psi})(\partial_{i}{\cal{X}})\Bigg]\Bigg\}\,.\hspace{0.8cm}
\end{eqnarray}

Now, we are in the position to calculate the three point correlation
function by using the interacting picture. In this picture, we
obtain the vacuum expectation value of the curvature perturbation
$\Psi$ for the three-point operator in the conformal time interval
between the beginning and end of the inflation
as~\cite{Mal03,Che08,See05}
\begin{eqnarray}
\label{eq38} \langle
{\Psi}(\textbf{k}_{1})\,{\Psi}(\textbf{k}_{2})\,{\Psi}(\textbf{k}_{3})\rangle
=-i\int_{\tau_{i}}^{\tau_{f}}d \tau \, a\, \langle0|
[{\Psi}(\tau_{f},\textbf{k}_{1})\,{\Psi}(\tau_{f},\textbf{k}_{2})\,{\Psi}(\tau_{f},\textbf{k}_{3}),
H_{int}(\tau)]|0\rangle,
\end{eqnarray}
where the interacting Hamiltonian, $H_{int}$, is equal to the
lagrangian of the third order action. To solve the integral of
equation \eqref{eq38}, we can approximate the coefficients in the
brackets of the lagrangian \eqref{eq37} to be constants since these
coefficients would vary slower than the scale factor. By solving the
integral, we find the three-point correlation function of the
curvature perturbation in the Fourier space as
\begin{eqnarray}
\label{eq39} \langle
{\Psi}(\textbf{k}_{1})\,{\Psi}(\textbf{k}_{2})\,{\Psi}(\textbf{k}_{3})\rangle
=(2\pi)^{3}\delta^{3}(\textbf{k}_{1}+\textbf{k}_{2}+\textbf{k}_{3}){\cal{B}}_{\Psi}(\textbf{k}_{1},\textbf{k}_{2},\textbf{k}_{3})\,,
\end{eqnarray}
where
\begin{equation}
\label{eq40}
{\cal{B}}_{\Psi}(\textbf{k}_{1},\textbf{k}_{2},\textbf{k}_{3})=\frac{(2\pi)^{4}{\cal{A}}_{s}}{\prod_{i=1}^{3}
k_{i}^{3}}\,
{\cal{E}}_{\Psi}(\textbf{k}_{1},\textbf{k}_{2},\textbf{k}_{3}).
\end{equation}
${\cal{A}}_{s}$ in equation \eqref{eq40} is given by equation
\eqref{eq21}. Also, the parameter ${\cal{E}}_{\Psi}$ is defined as
\begin{equation}
\label{eq41}
{\cal{E}}_{\Psi}=\frac{3}{4}\Bigg(1-\frac{1}{c_{s}^{2}}\Bigg){\cal{S}}_{1}+\frac{1}{4}\Bigg(1-\frac{1}{c_{s}^{2}}\Bigg){\cal{S}}_{2}
+\frac{3}{2}\Bigg(\frac{1}{c_{s}^{2}}-1\Bigg){\cal{S}}_{3},
\end{equation}
where
\begin{equation}
\label{eq42}
{\cal{S}}_{1}=\frac{2}{K}\sum_{i>j}k_{i}^{2}\,k_{j}^{2}-\frac{1}{K^{2}}\sum_{i\neq
j}k_{i}^{2}\,k_{j}^{3}\,,
\end{equation}
\begin{equation}
\label{eq43}
{\cal{S}}_{2}=\frac{1}{2}\sum_{i}k_{i}^{3}+\frac{2}{K}\sum_{i>j}k_{i}^{2}\,k_{j}^{2}-\frac{1}{K^{2}}\sum_{i\neq
j}k_{i}^{2}\,k_{j}^{3}\,,
\end{equation}
\begin{equation}
\label{eq44}
{\cal{S}}_{3}=\frac{\left(k_{1}\,k_{2}\,k_{3}\right)^{2}}{K^{3}}\,,
\end{equation}
and
\begin{equation}
\label{eq45} K=k_{1}+k_{2}+k_{3}\,.
\end{equation}
As it can be seen from equation \eqref{eq40}, a three-point
correlator depends on the three momenta $k_{1}$ , $k_{2}$ and
$k_{3}$. To satisfy the translation invariance, these momenta should
form a closed triangle, meaning that the sum of these momenta should
be zero ($k_{1}+k_{2}+k_{3}=0$). Also, satisfying the rotational
invariance makes the shape of the triangle
important~\cite{Bab04a,Kom05,Cre06,Lig06,Yad07}. Depending on the
values of momenta, there are several shapes and each shape has a
maximal signal in a special configuration of triangle. One of these
shapes is a local shape \cite{Gan94,Ver00,Wan00,Kom01} which has a
maximal signal in the squeezed limit ($k_{3}\ll k_{1}\simeq k_{2}$).
Another shape which is corresponding to the equilateral triangle
\cite{Bab04b}, has a signal at $k_{1}=k_{2}=k_{3}$. A linear
combination of the equilateral and orthogonal templates, which is
orthogonal \cite{Sen10} to equilateral one, gives a shape
corresponding to folded triangle \cite{Che07} with a pick in
$k_{1}=2k_{2}=2k_{3}$ limit. We mention that, the orthogonal
configuration has a signal with a positive peak at the equilateral
configuration and a negative peak at the folded configuration. In
order to measure the amplitude of the non-Gaussianity we define the
dimensionless parameter $f_{_{NL}}$, called ``nonlinearity
parameter'', as follows
\begin{equation}
\label{eq46}
f_{NL}=\frac{10}{3}\frac{{\cal{E}}_{\Psi}}{\sum_{i=1}^{3}k_{i}^{3}}\,.
\end{equation}
In this paper, we study the non-Gaussianity in the equilateral and
orthogonal configurations. To this end, we should find
${\cal{E}}_{\Psi}$ in these configurations. In this regard, we
should estimate the correlation between two different shapes. So,
following Refs.~\cite{Fer09,Fel13,Byr14}, we define the following
quantity
\begin{equation}
\label{eq46-2}
{\cal{C}}(\check{{\cal{B}}}_{\Psi}\check{{\cal{B}}}'_{\Psi})
=\frac{{\cal{F}}(\check{{\cal{B}}}_{\Psi}\check{{\cal{B}}}'_{\Psi})}{\sqrt{{\cal{F}}(\check{{\cal{B}}}_{\Psi}\check{{\cal{B}}}_{\Psi})
{\cal{F}}({\cal{B}}'_{\Psi}{\cal{B}}'_{\Psi})}}
\end{equation}
where,
$\check{{\cal{B}}}_{\Psi}=\frac{{\cal{B}}_{\Psi}}{{\cal{A}}_{s}}$
and
\begin{equation}
\label{eq46-2}
{\cal{F}}(\check{{\cal{B}}}_{\Psi}\check{{\cal{B}}}'_{\Psi})= \int
dk_{1}\,dk_{2}\,dk_{3}\,\check{{\cal{B}}}_{\Psi}(k_{1},k_{2},k_{3})\check{{\cal{B}}}'_{\Psi}(k_{1},k_{2},k_{3})
w
\end{equation}
with $w=\frac{(k_{1}k_{2}k_{3})^{4}}{(k_{1}+k_{2}+k_{3})^{3}}$. The
region of integration is $0<k_{1}<\infty$,
$0\leq\frac{k_{2}}{k_{1}}<1$ and
$-\frac{k_{2}}{k_{1}}\leq\frac{k_{3}}{k_{1}}\leq1$. We note that two
shapes satisfying the condition $\mid
{\cal{C}}(\check{{\cal{B}}}_{\Psi}\check{{\cal{B}}}'_{\Psi})
\mid\simeq 0$, are almost orthogonal. Now, we introduce a shape
$\breve{{\cal{S}}}^{equi}$ as~\cite{Fel13,Noz13c}
\begin{equation}
\label{eq47}
\breve{{\cal{S}}}^{equi}=-\frac{12}{13}\Big(3{\cal{S}}_{1}-{\cal{S}}_{2}\Big)\,.
\end{equation}
By using this equation and equation \eqref{eq46-2} we can show that
the follwing shape is orthogonal to \eqref{eq47}
\begin{equation}
\label{eq48}
\breve{{\cal{S}}}^{ortho}=\frac{12}{14-13\beta}\Big(\beta\big(3{\cal{S}}_{1}-{\cal{S}}_{2}\big)+3{\cal{S}}_{1}-{\cal{S}}_{2}\Big)\,,
\end{equation}
where $\beta\simeq 1.1967996$. Now, the bispectrum \eqref{eq41}, can
be expressed in terms of the equilateral and orthogonal basis as
\begin{equation}
\label{eq49}
{\cal{E}}_{\Psi}={\cal{C}}_{1}\,\breve{{\cal{S}}}^{equi} +
{\cal{C}}_{2} \,\breve{{\cal{S}}}^{ortho}\,,
\end{equation}
where the parameters ${\cal{C}}_{1}$ and ${\cal{C}}_{2}$ are given
by the following expressions
\begin{equation}
\label{eq50}
{\cal{C}}_{1}=\frac{13}{12}\Bigg[\frac{1}{24}\bigg(1-\frac{1}{c_{s}^{2}}\bigg)\bigg(2+3\beta\bigg)
\Bigg]\,,
\end{equation}
and
\begin{equation}
\label{eq51}
{\cal{C}}_{2}=\frac{14-13\beta}{12}\Bigg[\frac{1}{8}\bigg(1-\frac{1}{c_{s}^{2}}\bigg)\Bigg]\,.
\end{equation}
By using equations \eqref{eq46}-\eqref{eq51}, we can obtain the
amplitude of the non-Gaussianity in the equilateral and orthogonal
configuration as follows
\begin{equation}
\label{eq52}
f_{_{NL}}^{equi}=\frac{130}{36\sum_{i=1}^{3}k_{i}^{3}}\Bigg[\frac{1}{24}\bigg(1-\frac{1}{c_{s}^{2}}\bigg)\bigg(2+3\beta\bigg)
\Bigg]\breve{{\cal{S}}}^{equi}\,,
\end{equation}
and
\begin{equation}
\label{eq53}
f_{_{NL}}^{ortho}=\frac{140-130\beta}{36\,\sum_{i=1}^{3}k_{i}^{3}}\Bigg[\frac{1}{8}\bigg(1-\frac{1}{c_{s}^{2}}\bigg)
\Bigg]\breve{{\cal{S}}}^{ortho}\,.
\end{equation}
Since the equilateral shape has a maximal signal at
$k_{1}=k_{2}=k_{3}$ limit and also, an orthogonal configuration has
a positive peak at this limit, so we rewrite equations \eqref{eq52}
and \eqref{eq53} in this limit as
\begin{equation}
\label{eq54}
f_{_{NL}}^{equi}=\frac{325}{18}\Bigg[\frac{1}{24}\bigg(\frac{1}{c_{s}^{2}}-1\bigg)\bigg(2+3\beta\bigg)
\Bigg]\,,
\end{equation}
and
\begin{equation}
\label{eq55}
f_{_{NL}}^{ortho}=\frac{10}{9}\Big(\frac{65}{4}\beta+\frac{7}{6}\Big)\Bigg[\frac{1}{8}\bigg(1-\frac{1}{c_{s}^{2}}\bigg)
\Bigg]\,.
\end{equation}

So far, we have obtained the main equations of our setup. In the
following section, we examine our model by using the recently
released observational data by Planck Collaboration and find some
constraints on the model's parameters space.

\section{Observational Constraints}

In this section we are going to find some observational constraints
on the parameters space of the model in hand. To this end, first we
define the extension of the nonminimal derivative term as
${\cal{F(\phi)}}\sim\frac{1}{2n}\phi^n$, which for $n=1$ leads to
the usual nonminimal derivative case (as mentioned in the
Introduction). We set the values $1$, $2$, $3$ and $4$ for $n$.
Another general function in this setup is $f(\phi)$, which shows the
nonminimal coupling between the scalar field and the curvature
scalar. For this function we consider two cases: $f(\phi)\sim
\xi\phi^2$ and $f(\phi)\sim \xi\phi^4$. The last general function is
the potential term in the action. We adopt various types of
potential to explore: the linear potential as
$V(\phi)\sim\phi$~\cite{All10}, the quadratic potential as
$V(\phi)\sim\phi^2$, the cubic potential as $V(\phi)\sim\phi^3$, the
quartic potential as $V(\phi)\sim\phi^4$, the axion monodromy's
motivated potential as $V(\phi)\sim\phi^{\frac{2}{3}}$~\cite{Sil08}
and the exponential potential as $V(\phi)\sim e^{-\kappa\phi}$. With
this specification of the general functions, we can analyze the
model numerically and find some constraints on the nonminimal
coupling parameter $\xi$. For this purpose, we solve the integral of
equation \eqref{eq8} by adopting the mentioned functions of the
scalar field. By solving this equation, we find the value of the
inflaton at the horizon crossing in terms of the e-folds number. By
substituting $\phi_{hc}$ in equations \eqref{eq23}, \eqref{eq34},
\eqref{eq54} and \eqref{eq55}, we can find the scalar spectral
index, tensor-to-scalar ratio, the amplitude of the orthogonal and
equilateral configuration of the non-Gaussianity in terms of the
e-folds number. Then we can analyze these parameters numerically to
see the observational viability of this setup in confrontation with
recently released data. In plotting the figures and analyzing the
model we have re-scaled all constant parameter to unity with the
choice $G=c=1$. In this regard, we study the behavior of the
tensor-to-scalar ratio versus the scalar spectral index and the
orthogonal configuration versus the equilateral configuration. The
observational parameters are defined at $k_{0}=0.002 Mpc^{-1}$ where
subscript $0$ refers to the value of $k$ when it left the Hubble
radius during inflation. The results are shown in figures 1-8. We
have found that an inflationary model with a nonminimal coupling
between the scalar filed and Ricci scalar and also a nonminimal
derivative coupling, in some ranges of nonminimal coupling parameter
$\xi$ is consistent with Planck2015 dataset. In fact, these
nonminimal couplings help to make models of chaotic inflation, that
would otherwise be in tension with Planck data, in better agreement
with the data. The ranges of nonminimal coupling are shown in tables
I-IV. Note that, some other authors have considered inflationary
model with nonminimal coupling term and studied their model
numerically. We can compare the result of our setting with their
results here. For example, it is shown in Ref.~\cite{Lee14} that the
presence of nonminimal coupling and non-canonical kinetic terms lead
to an chaotic inflation model, which is favored by the BICEP2
observation but is in tension with recent observational data. Here,
we have shown that the presence of nonminimal coupling between the
scalar field and Ricci scalar and also a general nonminimal
derivative coupling between the scalar field and gravity make the
model observationally more viable. This model with these couplings
is consistent with Planck2015 TT, TE, EE+lowP data. The authors of
Ref.~\cite{Gla15} have considered a nonminimal coupling between a
scalar field and gravity as $(\alpha\phi^{2}+\beta\phi^{4})R$ in
Jordan frame. They perform a frame transformation to Einstein frame
and compare their model with Planck2015 data. Their model just for
$N=65$ and with some maximum values of $n_{s}$ is consistent with
observational data and for $N=50$ and $60$ is observationally
disfavored. In our setup, we have considered nonminimal coupling
function $\phi^{2}$ and $\phi^{4}$ separately. Our numerical
analysis shows that the presence of NMC and NMDC makes the model
observationally viable for both $N=50$ and $N=60$ in  the Jordan
frame (note that although we don't present here the results with
$N=50$, but we have analyzed the model with this value of number of
e-folds and have found the viability of the model). In
Ref.~\cite{Bou15} a nonminimally coupled inflaionary model with
quadratic potential in Einstein frame is considered in the
background of Planck2015 observational data. Their numerical
analysis shows that this model for some value of NMC parameter gives
$n_{s}\simeq 0.9652$. Interestingly our setup also, in some values
of $\xi$ gives $n_{s}\simeq 0.9652$. Nevertheless, we should
emphasize that for some other choices for model parameters, this
would be not the case necessarily.

Our numerical analysis shows that this nonminimal model with small
values of the NMC parameter is consistent with observational data
and so this model in Jordan frame in weak coupling limit has
cosmological viability. In week coupling limit our setup is
applicable to several inflationary potentials. In this limit and in
the ranges obtained by numerical analysis, the scalar spectral index
is red tilted. Another point is that in our setup, in some ranges of
NMC parameter, it is possible to have large non-Gaussianity, as can
be seen in figures. Note that, to obtain the constraints on $\xi$,
we have focused on those ranges of the NMC parameter that lead to
the values of the observable quantities allowed by the Planck 2015
observational data. For instance, we have focused on the values of
$\xi$ leading to $0.9522< n_{s}<0.977$ allowed by Planck2015 TT, TE,
EE+lowP data. In the same way, we focused on the values of the NMC
parameter leading to $-147<f_{NL}^{equi}<143$ allowed by Planck2015
TTT, EEE, TTE and EET data. Also, equations \eqref{eq23}),
\eqref{eq34}, \eqref{eq54} and \eqref{eq55} show that the scalar
spectral index, tensor-to-scalar ratio and the amplitudes of
non-Gaussianity have nonlinear dependence on the NMC parameter
$\xi$. The function $f(\phi)$ has an explicit non-minimal coupling
dependence. In the mentioned equations, there are different powers
of the function $f(\phi)$ and therefore different powers of the
nonminimal coupling parameter. Because of complexity of equations we
have avoided to rewrite the equations in terms of the NMC parameter
$\xi$. However, we note that the effects of these nonlinearities in
$\xi$ have been included in our analysis and graphs. Since
quantities such as the scalar spectral index, tensor-to-scalar ratio
and the amplitudes of non-Gaussianity have nonlinear dependence on
the NMC parameter $\xi$, any small change in the values of $\xi$ has
considerable effect on the values of these quantities. In fact,
there is a nonlinear dependence on model parameters that really
demands four figures of accuracy in quantities such as $\xi$ to get
two figures of accuracy in observable quantities such as $n_{s}$.
So, we have tried to be more accurate and found the values of $\xi$
with higher precision.

\begin{figure*}
\flushleft\leftskip3em{
\includegraphics[width=.38\textwidth,origin=c,angle=0]{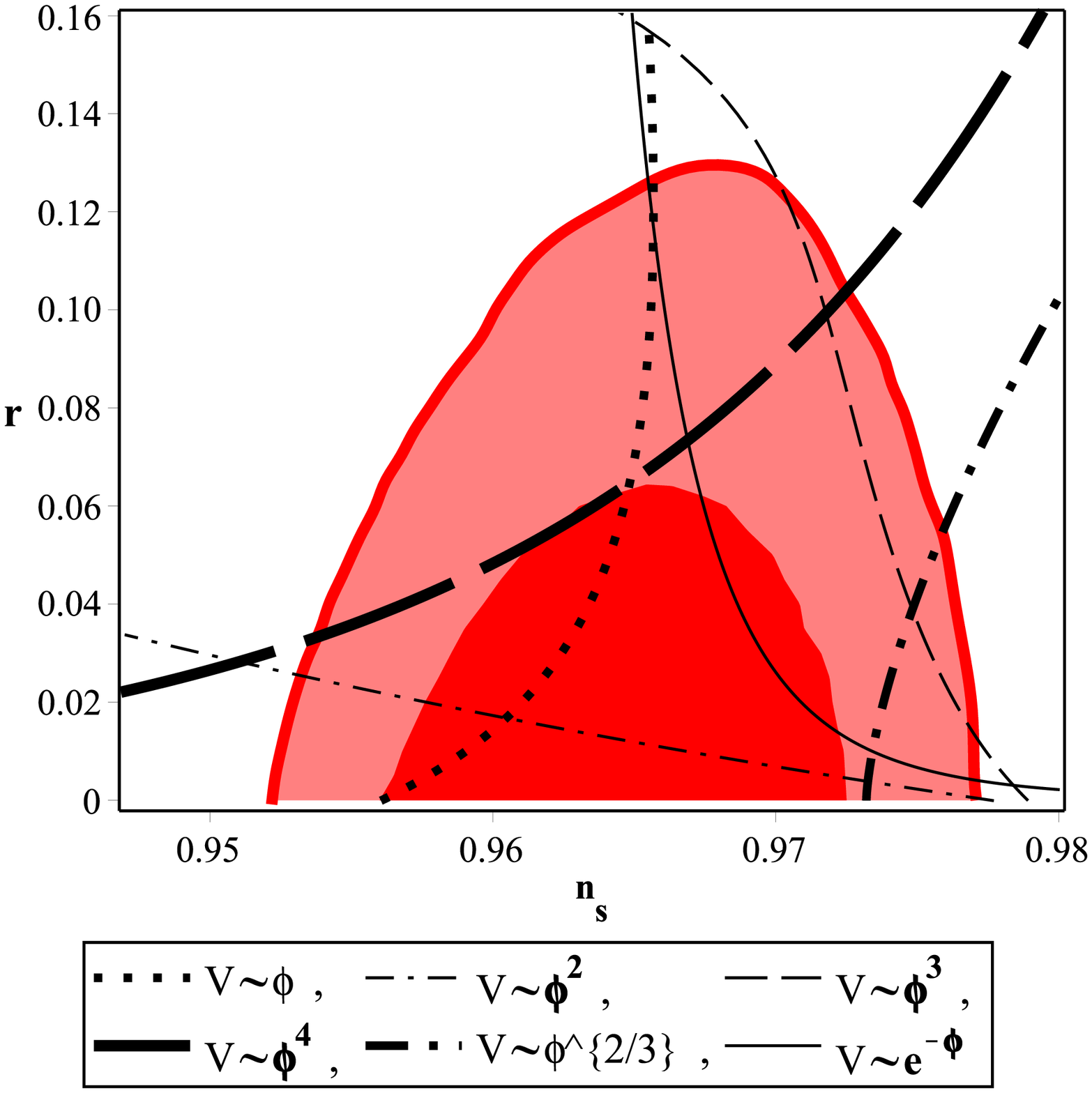}
\hspace{1cm}
\includegraphics[width=.38\textwidth,origin=c,angle=0]{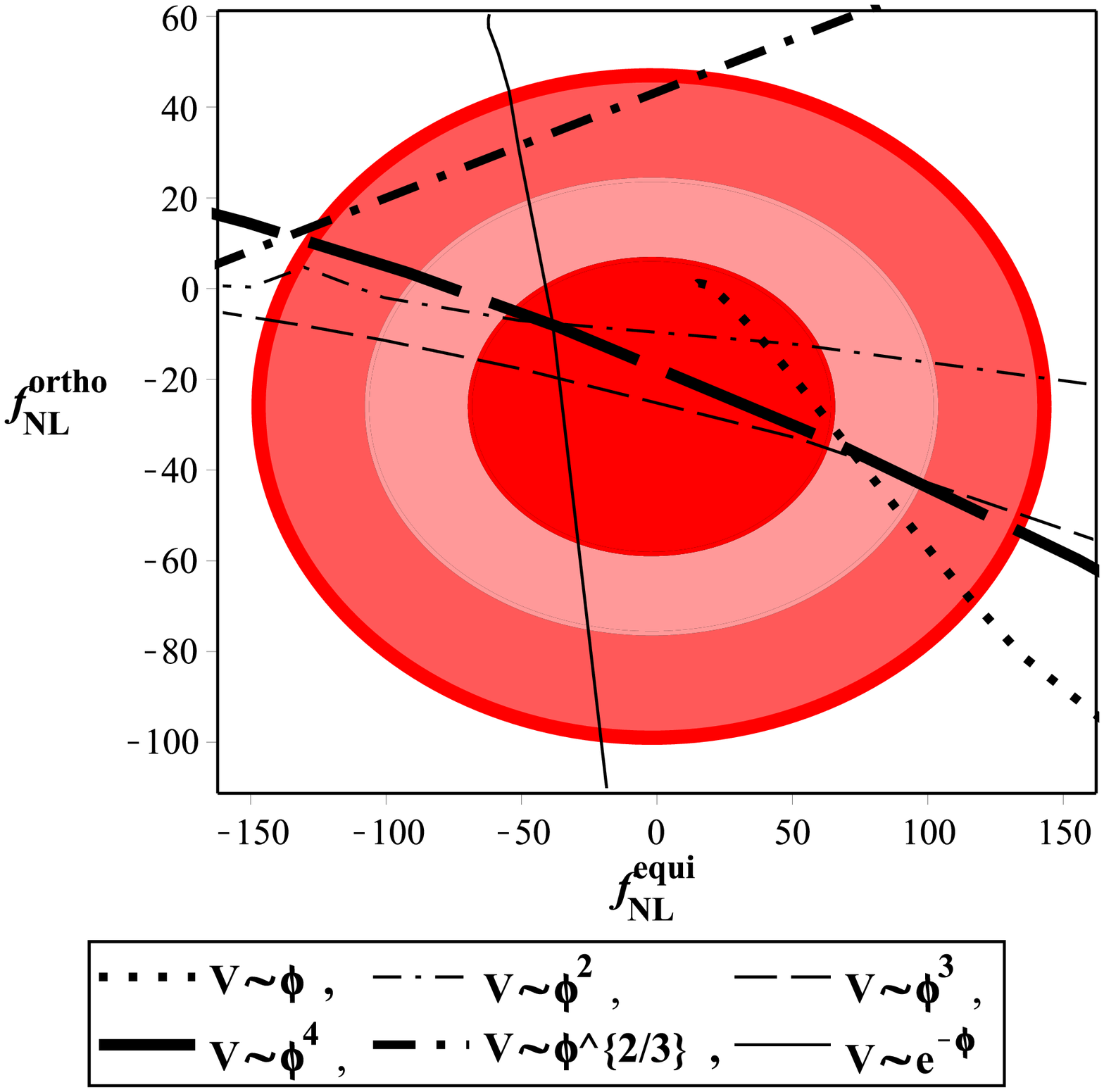}}
\caption{\label{fig1} Tensor-to-scalar ratio versus the scalar
spectral index for an inflationary model in which both the inflaton
and its derivative are nonminimally coupled to gravity, in the
background of the Planck2015 TT, TE, EE+lowP data (left panel) and
the amplitude of the orthogonal configuration of the non-Gaussianity
versus the amplitude of the equilateral configuration for an
inflationary model in which both inflaton and its derivative are
nonminimally coupled to gravity, in the background of Planck2015
TTT, EEE, TTE and EET data (right panel). Figure is plotted with
$n=1$, $f(\phi)\sim \xi\phi^2$ and various types of the potential,
for $N=60$}
\end{figure*}
\begin{table*}
\begin{tiny}
\begin{center}
\caption{\label{tab:1} The ranges of $\xi$ in which the values of
the inflationary parameters $r$ and $n_{s}$ and also,
$f_{NL}^{ortho}$ and $f_{NL}^{equi}$, with $n=1$, are compatible
with the $95\%$ CL of the Planck2015 dataset.}
\begin{tabular}{ccccc}
\\ \hline \hline$V$&  $r$ - $n_{s}$ &$r$ - $n_{s}$&
$f_{NL}^{ortho}$ - $f_{NL}^{equi}$ &$f_{NL}^{ortho}$ - $f_{NL}^{equi}$ \\
\hline  & $f(\phi)\sim \xi\phi^2$ & $f(\phi)\sim \xi\phi^4$& $f(\phi)\sim \xi\phi^2$ & $f(\phi)\sim \xi\phi^4$ \\
\hline\\ $\phi$& $\xi<0.091$ & $\xi<0.0912$&$\xi<0.090$
& $0.008<\xi<0.091$\\\\
$\phi^{2}$&   $\xi<0.088$
 & $\xi<0.09$ &$0.01<\xi<0.089$ &$0.015<\xi<0.0895$\\\\
$\phi^{3}$&   $0.012<\xi<0.09$
 & $\xi<0.08$ &$0.011<\xi<0.093$&$0.021<\xi<0.082$\\\\
$\phi^{4}$&  $0.014<\xi<0.10$
 & $0.001<\xi<0.093$&$0.013<\xi<0.10$ &$0.01<\xi<0.098$\\\\
$\phi^{\frac{2}{3}}$&  $\xi<0.081$
 & $\xi<0.082$ & $0.015<\xi<0.073$&$0.01<\xi<0.109$\\\\
$e^{-\phi}$&  $0.016<\xi<0.108$
 & $0.078<\xi<0.080$ & $0.094<\xi<0.103$&$0.017<\xi<0.097$ \\\\
\hline\\\\
\end{tabular}
\end{center}
\end{tiny}
\end{table*}
\begin{figure*}
\flushleft\leftskip3em{
\includegraphics[width=.38\textwidth,origin=c,angle=0]{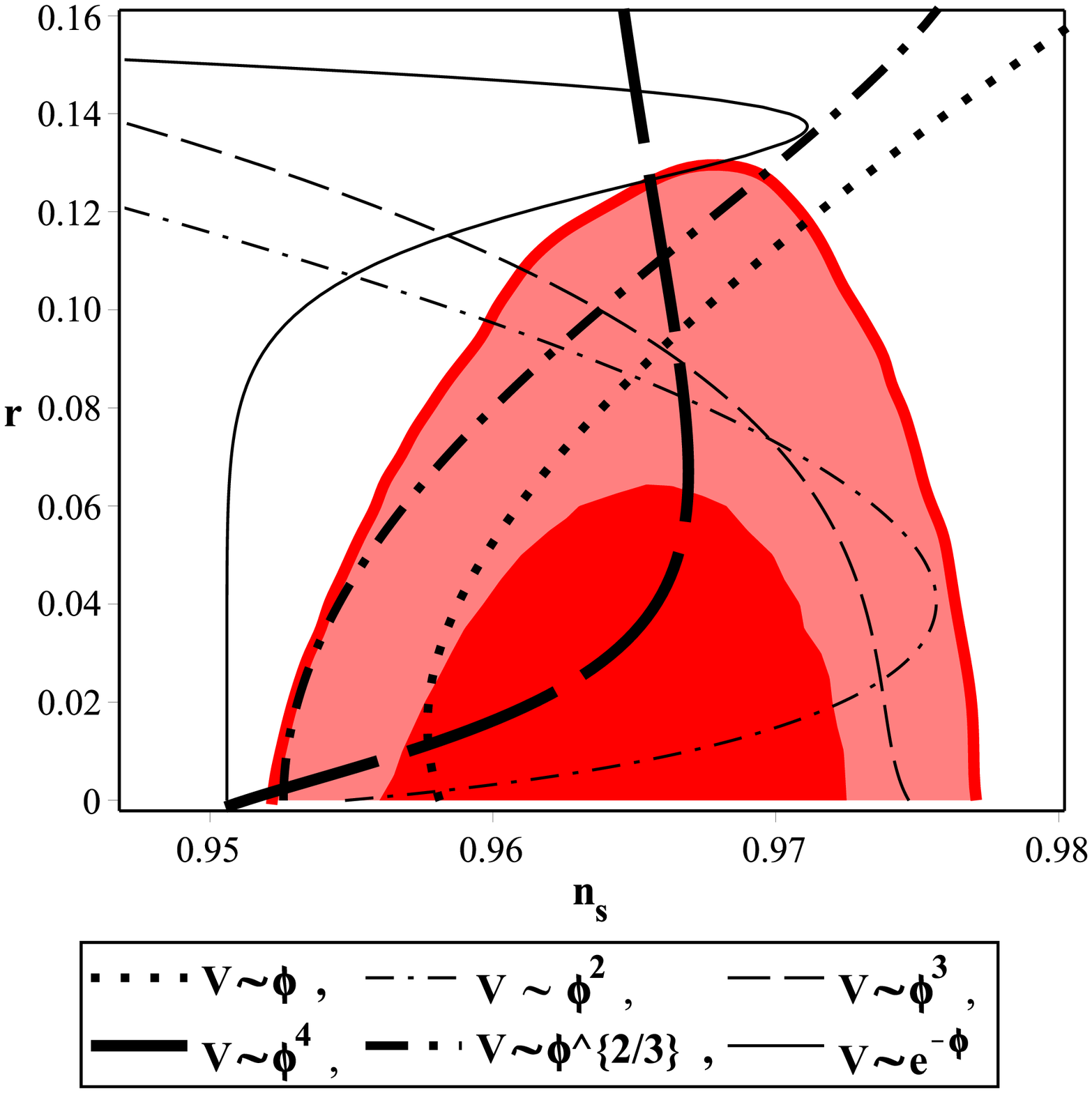}
\hspace{1cm}
\includegraphics[width=.38\textwidth,origin=c,angle=0]{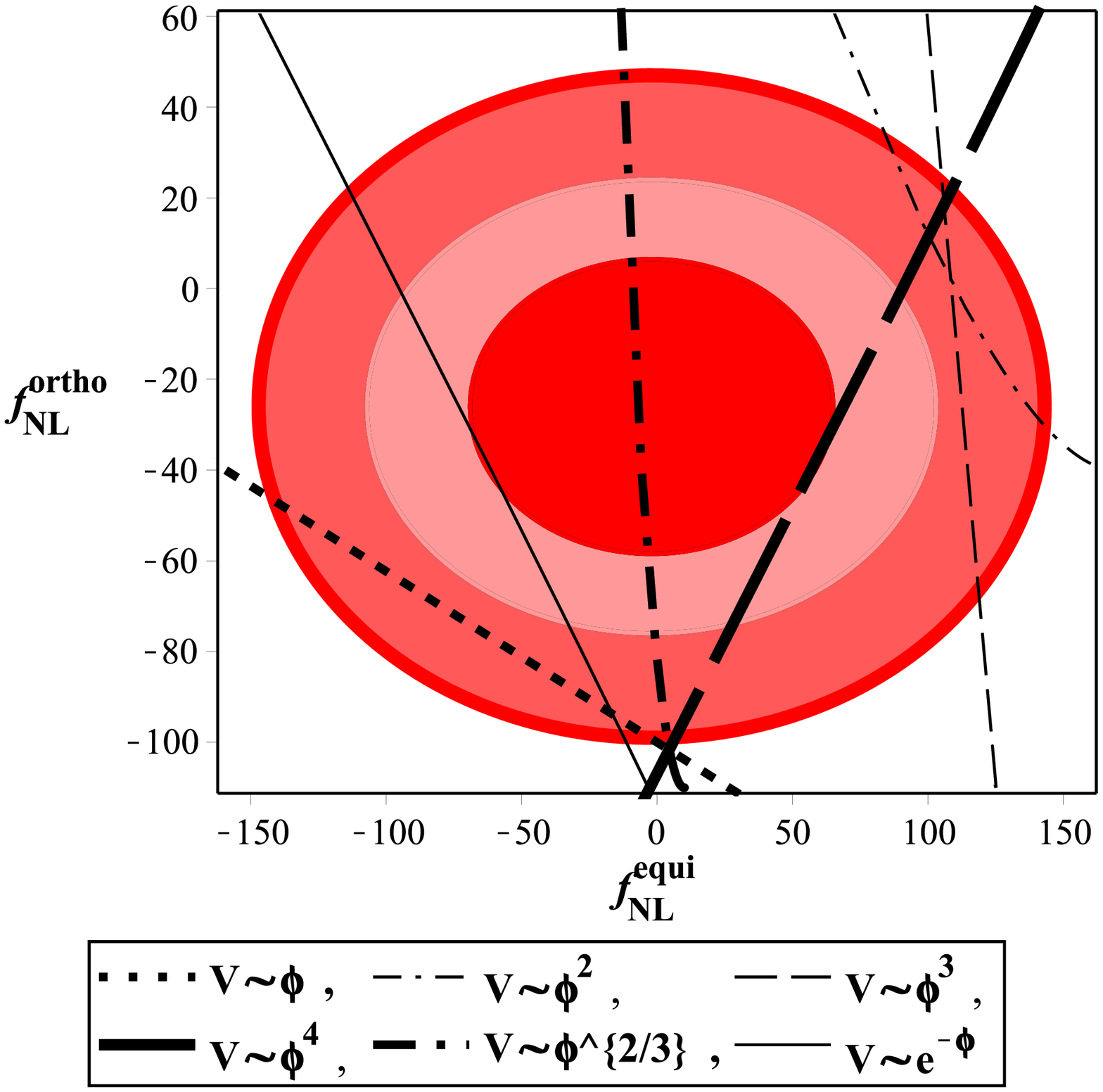}}
\caption{\label{fig3} Tensor-to-scalar ratio versus the scalar
spectral index for an inflationary model in which both inflaton and
its derivative are nonminimally coupled to gravity, in the
background of the Planck2015 TT, TE, EE+lowP data (left panel) and
the amplitude of the orthogonal configuration of non-Gaussianity
versus the equilateral configuration for an inflationary model in
which both inflaton and its derivative are nonminimally coupled to
gravity, in the background of Planck2015 TTT, EEE, TTE and EET data.
(right panel). Figure is plotted with $n=1$, $f(\phi)\sim \xi\phi^4$
and various types of potential, for $N=60$.}
\end{figure*}
\begin{figure*}
\flushleft\leftskip3em{
\includegraphics[width=.38\textwidth,origin=c,angle=0]{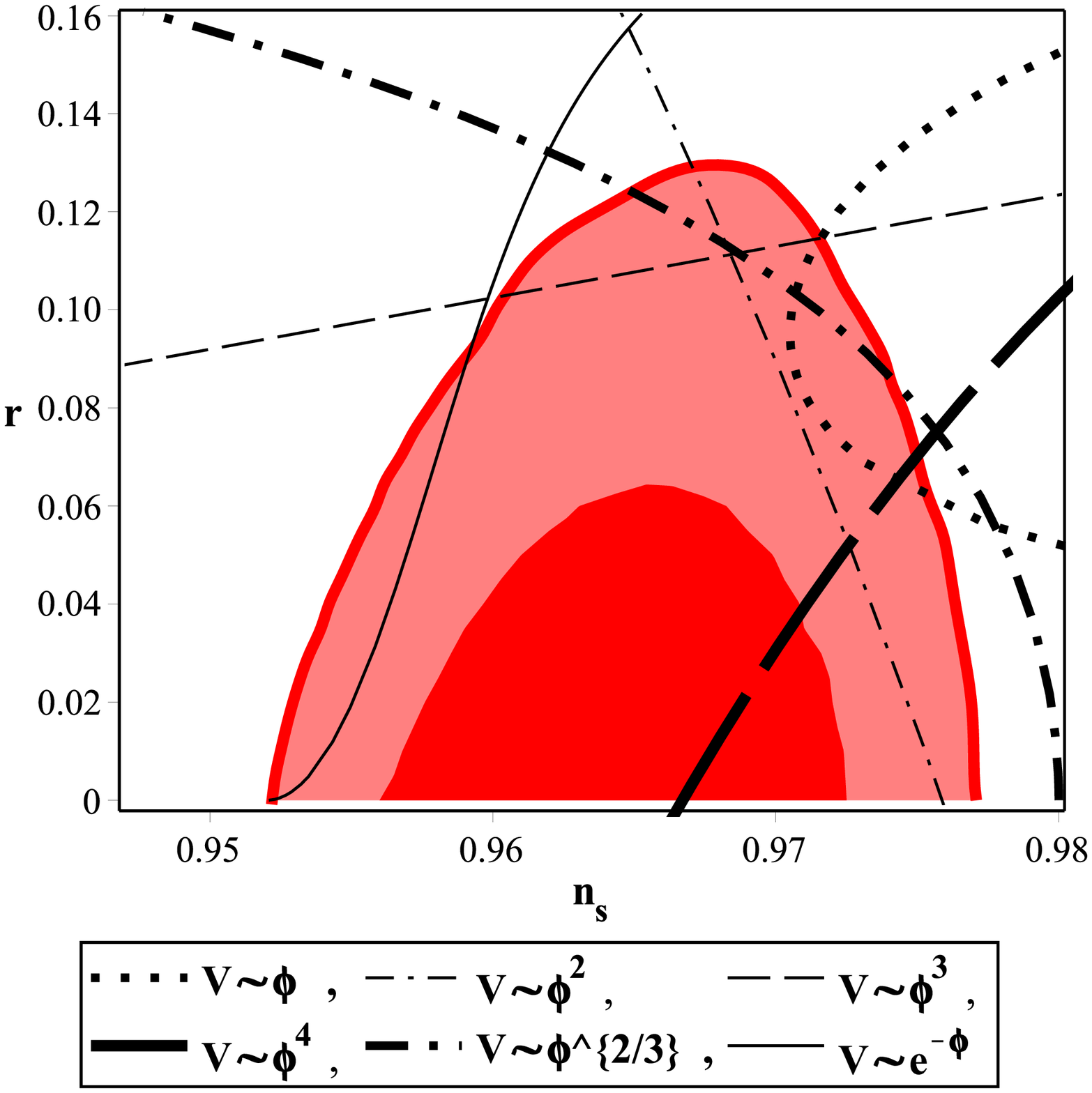}
\hspace{1cm}
\includegraphics[width=.38\textwidth,origin=c,angle=0]{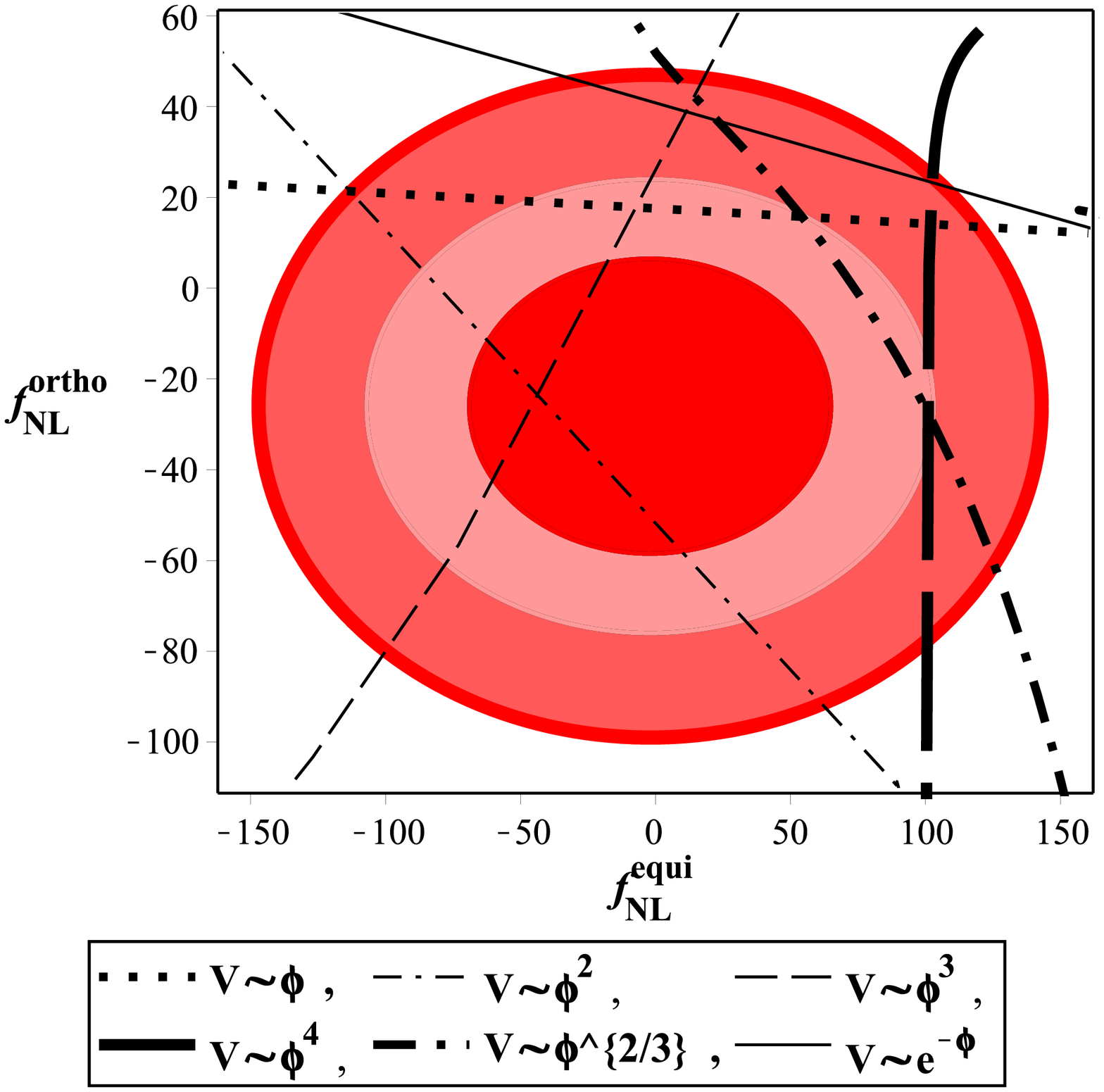}}
\caption{\label{fig5} Tensor-to-scalar ratio versus the scalar
spectral index for an inflationary model in which both inflaton and
its derivative are nonminimally coupled to gravity, in the
background of Planck2015 TT, TE, EE+lowP data (left panel) and the
amplitude of the orthogonal configuration versus the equilateral
configuration of non-Gaussianity for an inflationary model in which
both inflaton and its derivative are nonminimally coupled to
gravity, in the background of Planck2015 TTT, EEE, TTE and EET data
(right panel). Figure is plotted with $n=2$, $f(\phi)\sim \xi\phi^2$
and various types of potential, for $N=60$. }
\end{figure*}
\begin{table*}
\begin{tiny}
\begin{center}
\caption{\label{tab:2} The ranges of $\xi$ in which the values of
the inflationary parameters $r$ and $n_{s}$ and also,
$f_{NL}^{ortho}$ and $f_{NL}^{equi}$, with $n=2$, are compatible
with the $95\%$ CL of the Planck2015 dataset.}
\begin{tabular}{ccccc}
\\ \hline \hline$V$& $r$ - $n_{s}$ &$r$ - $n_{s}$&
$f_{NL}^{ortho}$ - $f_{NL}^{equi}$ & $f_{NL}^{ortho}$ - $f_{NL}^{equi}$ \\
\hline  & $f(\phi)\sim \xi\phi^2$ & $f(\phi)\sim \xi\phi^4$& $f(\phi)\sim \xi\phi^2$ & $f(\phi)\sim \xi\phi^4$ \\
\hline\\ $\phi$& $0.021<\xi<0.090$
& $0.019<\xi<0.09$ &$0.020<\xi<0.088$ &$0.021<\xi<0.097$\\\\
$\phi^{2}$&   $\xi<0.109$
 & $0.01<\xi<0.055$\,,\,$0.06<\xi<0.114$& $0.017<\xi<0.89$&$0.014<\xi<0.109$ \\\\
$\phi^{3}$&   $0.022<\xi<0.081$
 & $0.009<\xi<0.086$ & $0.018<\xi<0.098$&$\xi<0.086$\\\\
$\phi^{4}$&   $\xi<0.098$
 & $0.010<\xi<0.09$& $0.011<\xi<0.107$ &$0.0108<\xi<0.098$\\\\
$\phi^{\frac{2}{3}}$  & $0.021<\xi<0.076$
 & $0.018<\xi<0.084$&$0.020<\xi<0.086$ &$0.011<\xi<0.084$\\\\
$e^{-\phi}$&   $\xi<0.086$
 & $0.01<\xi<0.081$ & $0.028<\xi<0.089$ &$0.01<\xi<0.104$\\\\
\hline\\\\
\end{tabular}
\end{center}
\end{tiny}
\end{table*}
\begin{figure*}
\flushleft\leftskip3em{
\includegraphics[width=.38\textwidth,origin=c,angle=0]{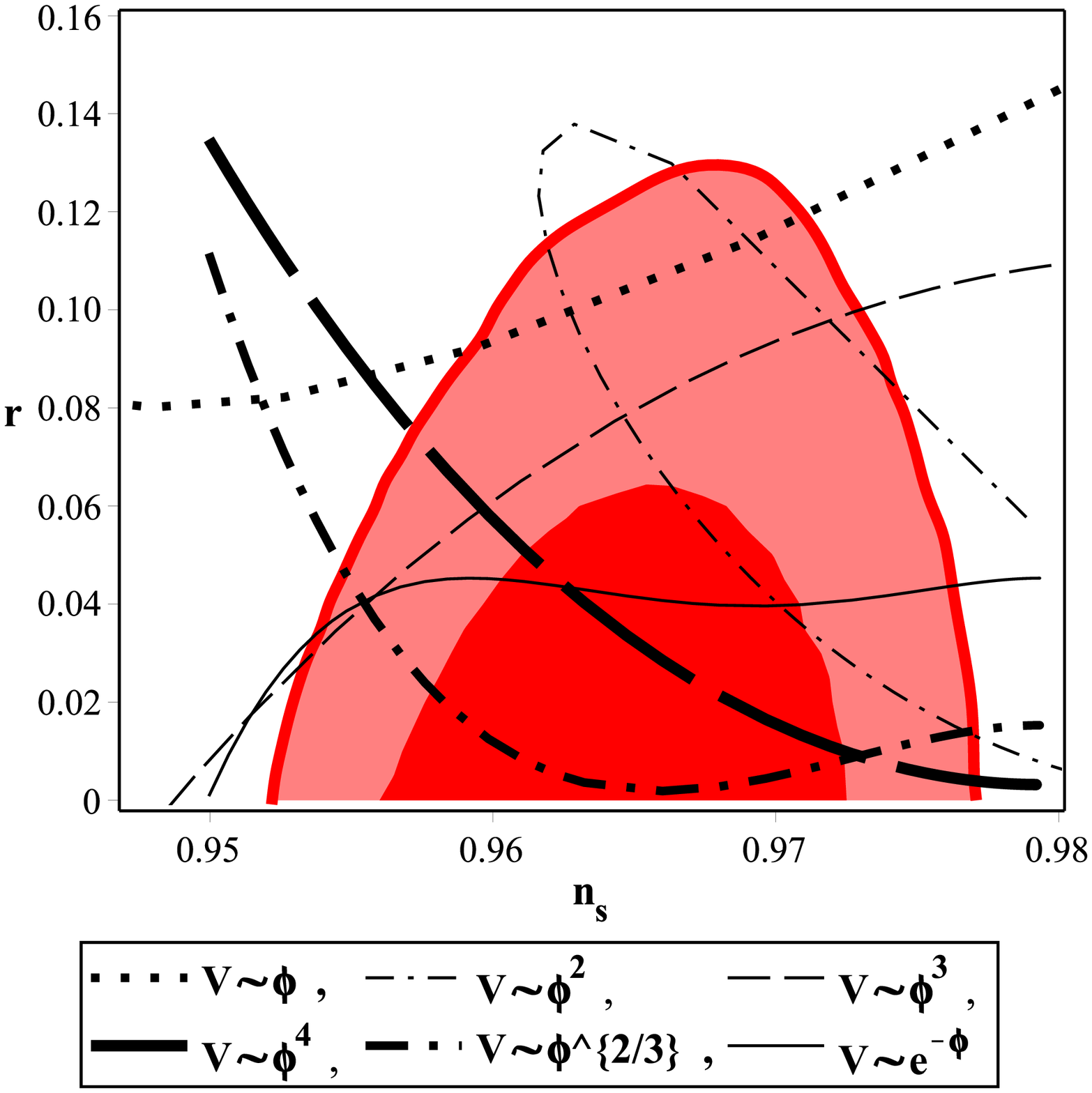}
\hspace{1cm}
\includegraphics[width=.38\textwidth,origin=c,angle=0]{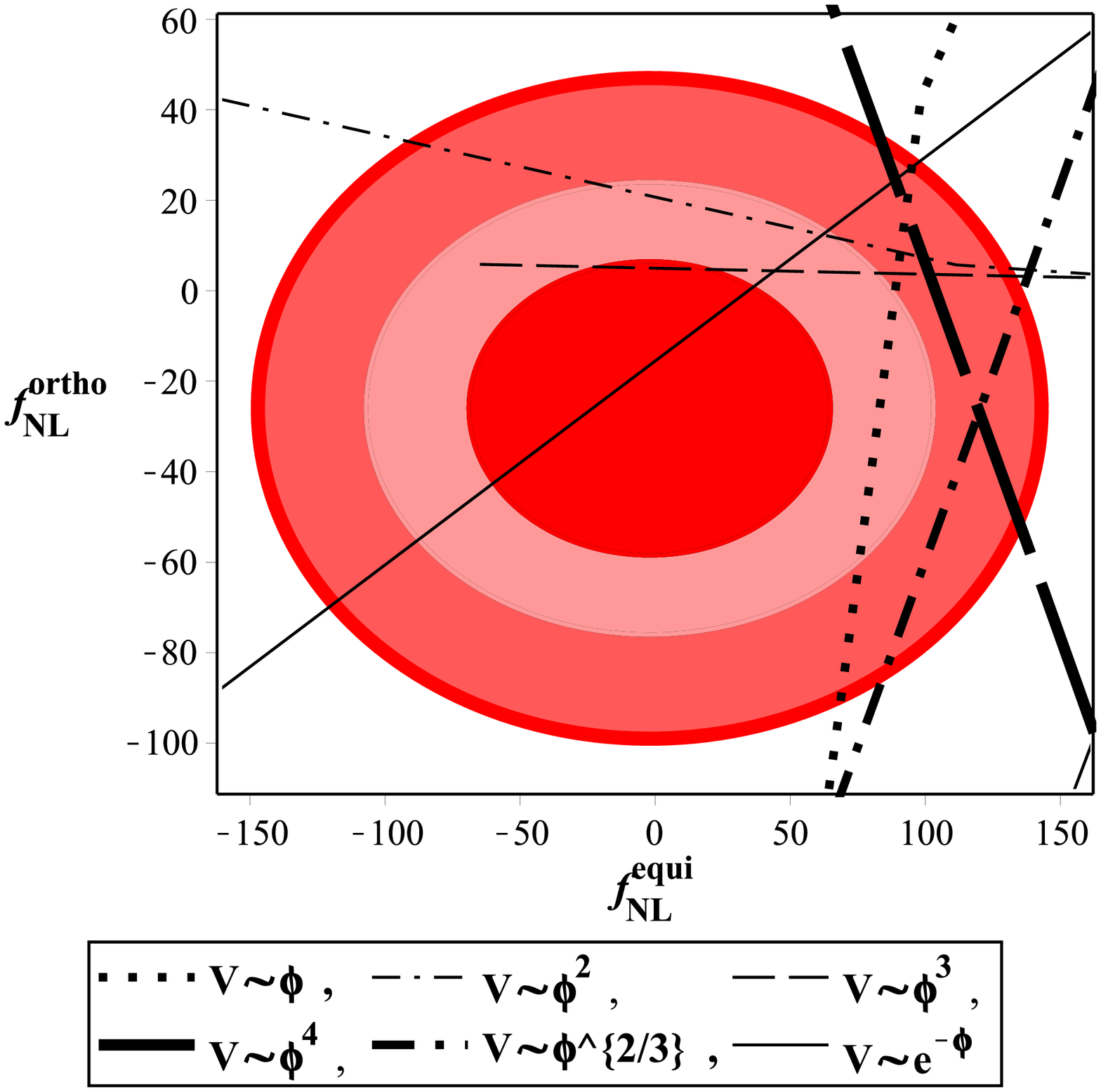}}
\caption{\label{fig7} Tensor-to-scalar ratio versus the scalar
spectral index for an inflationary model in which both inflaton and
its derivative are nonminimally coupled to gravity, in the
background of Planck2015 TT, TE, EE+lowP data (left panel) and tThe
amplitude of the orthogonal configuration versus the amplitude of
the equilateral configuration of non-Gaussianity for an inflationary
model in which both inflaton and its derivative are nonminimally
coupled to gravity, in the background of Planck2015 TTT, EEE, TTE
and EET data (right panel). Figure is plotted with $n=2$,
$f(\phi)\sim \xi\phi^4$ and various types of potential, for $N=60$.}
\end{figure*}
\begin{figure*}
\flushleft\leftskip3em{
\includegraphics[width=.38\textwidth,origin=c,angle=0]{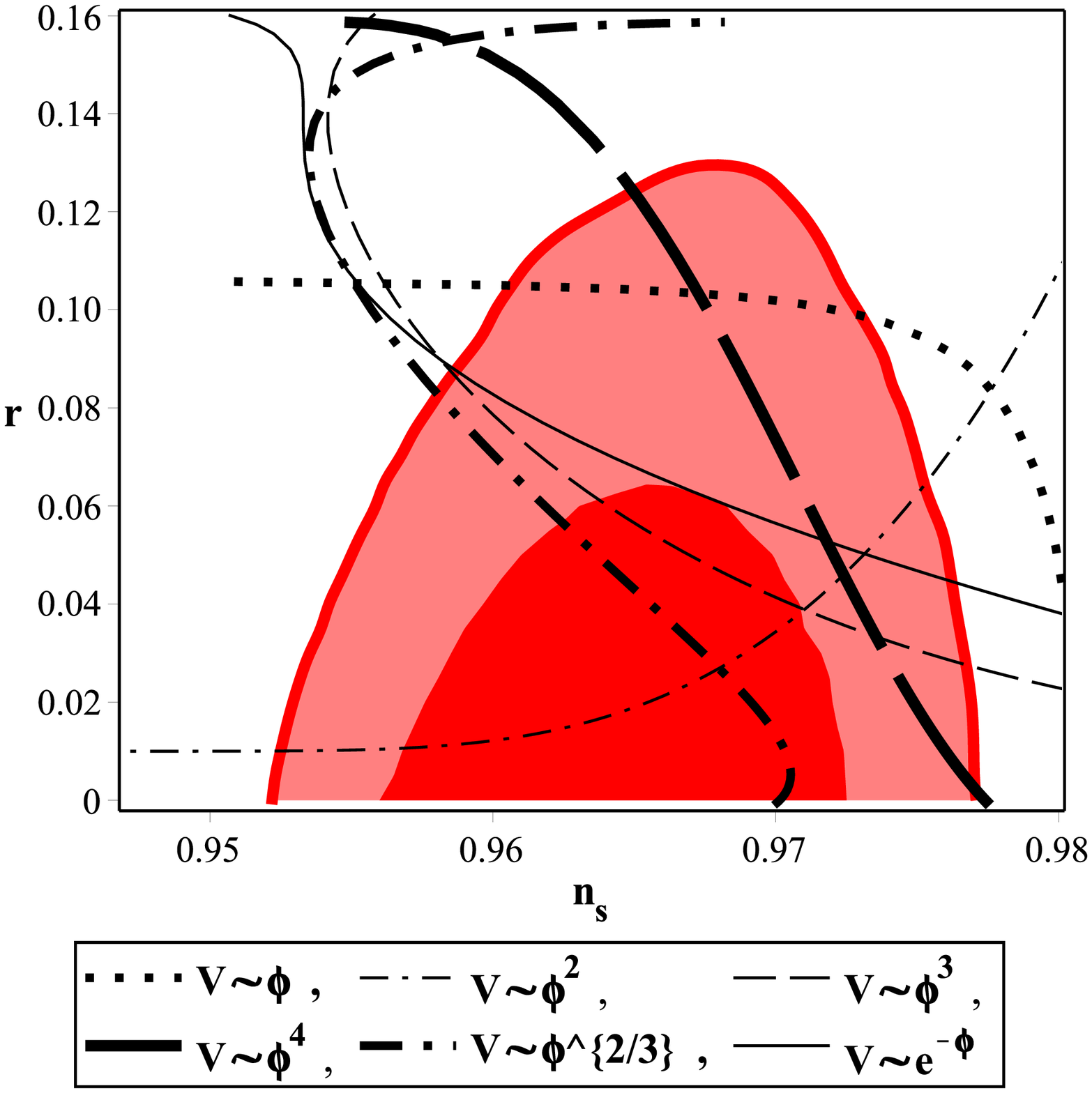}
\hspace{1cm}
\includegraphics[width=.38\textwidth,origin=c,angle=0]{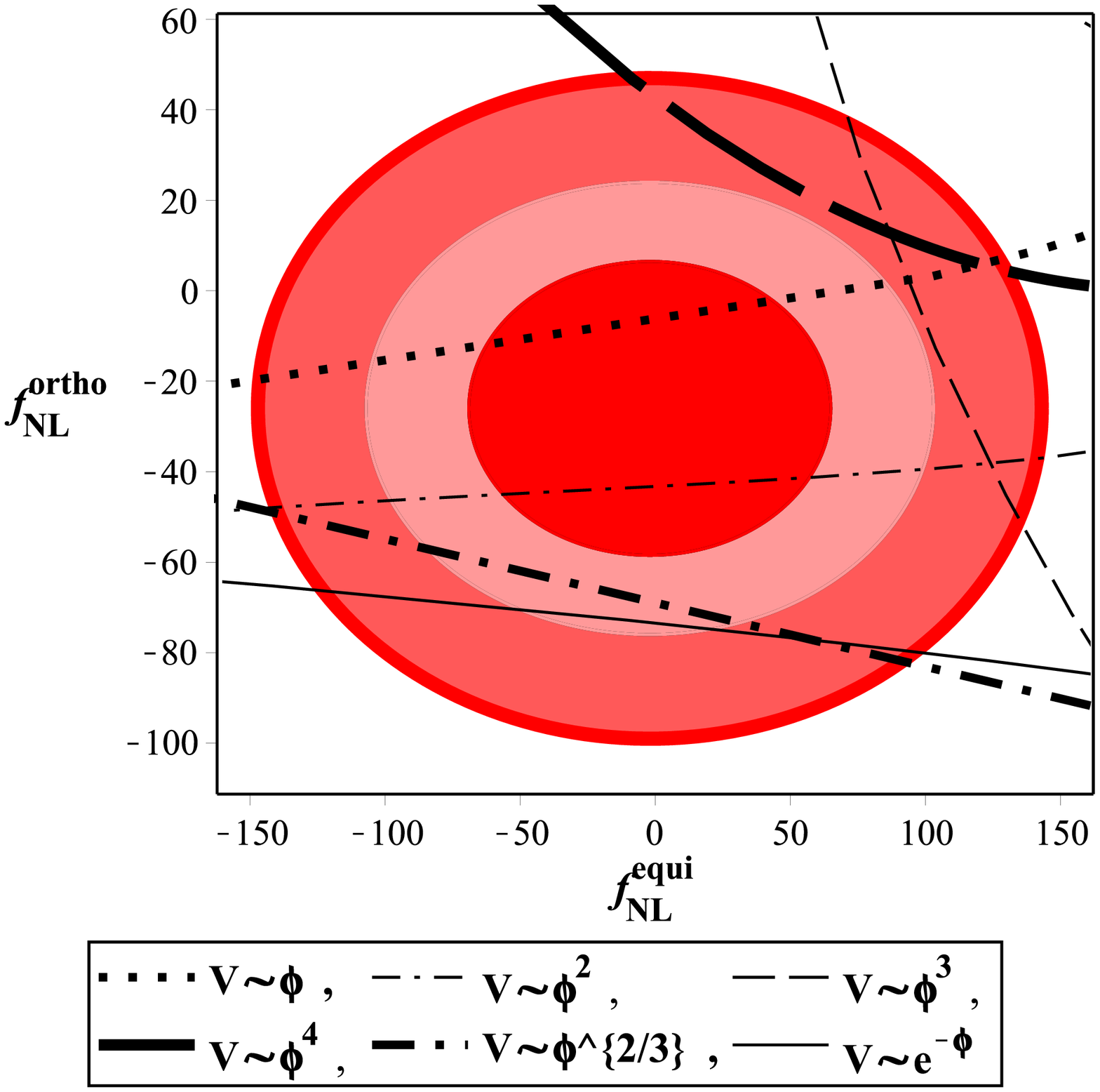}}
\caption{\label{fig9} Tensor-to-scalar ratio versus the scalar
spectral index for an inflationary model in which both inflaton and
its derivative are nonminimally coupled to gravity, in the
background of Planck2015 TT, TE, EE+lowP data (left panel) and the
amplitude of the orthogonal configuration versus the amplitude of
the equilateral configuration of non-Gaussianity for an inflationary
model in which both inflaton and its derivative are nonminimally
coupled to gravity, in the background of Planck2015 TTT, EEE, TTE
and EET data (right panel). Figure is plotted with $n=3$,
$f(\phi)\sim \xi\phi^2$ and various types of potential, for $N=60$.}
\end{figure*}
\begin{figure*}
\flushleft\leftskip3em{
\includegraphics[width=.38\textwidth,origin=c,angle=0]{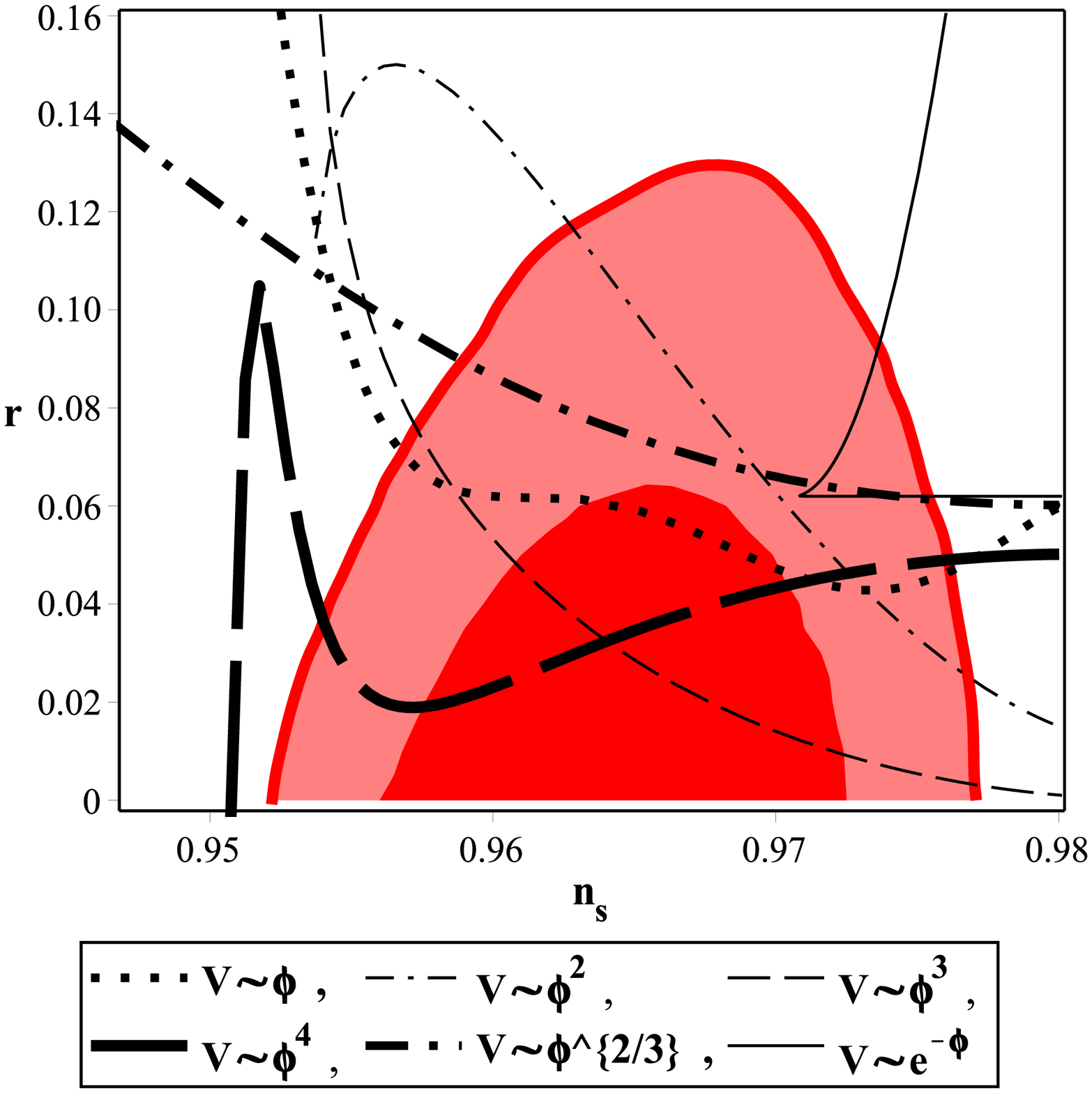}
\hspace{1cm}
\includegraphics[width=.38\textwidth,origin=c,angle=0]{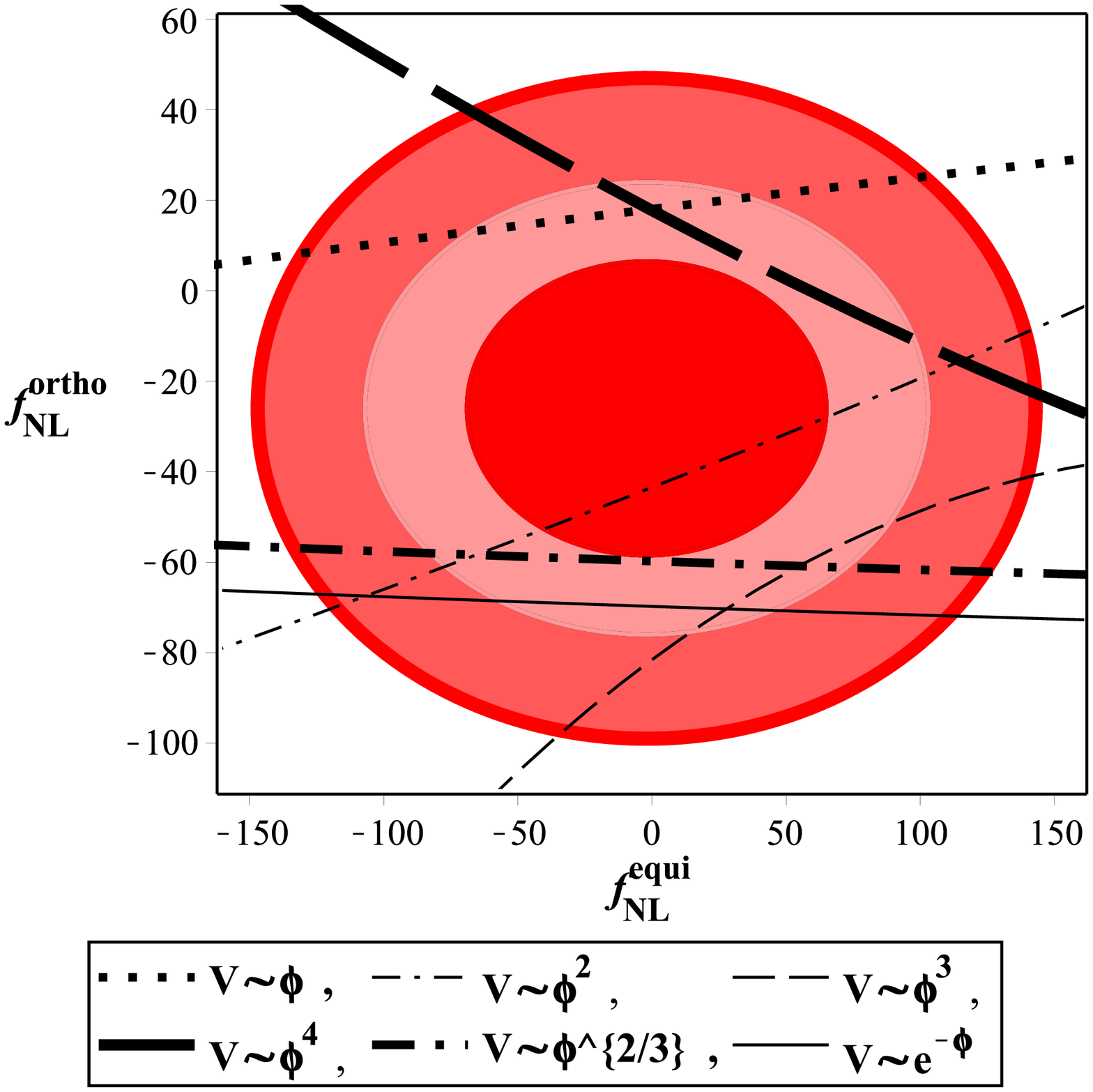}}
\caption{\label{fig11} Tensor-to-scalar ratio versus the scalar
spectral index for an inflationary model in which both inflaton and
its derivative are nonminimally coupled to gravity, in the
background of Planck2015 TT, TE, EE+lowP data (left panel) and the
amplitude of the orthogonal configuration versus the amplitude of
the equilateral configuration of non-Gaussianity for an inflationary
model in which both inflaton and its derivative are nonminimally
coupled to gravity, in the background of Planck2015 TTT, EEE, TTE
and EET data (right panel). Figure is plotted with $n=3$,
$f(\phi)\sim \xi\phi^4$ and various types of potential, for $N=60$.}
\end{figure*}
\begin{table*}
\begin{tiny}
\begin{center}
\caption{\label{tab:3} The ranges of $\xi$ in which the values of
the inflationary parameters $r$ and $n_{s}$ and also,
$f_{NL}^{ortho}$ and $f_{NL}^{equi}$, with $n=3$, are compatible
with the $95\%$ CL of the Planck2015 dataset.}
\begin{tabular}{ccccc}
\\ \hline \hline$V$& $r$ - $n_{s}$ &$r$ - $n_{s}$&
$f_{NL}^{ortho}$ - $f_{NL}^{equi}$ & $f_{NL}^{ortho}$ - $f_{NL}^{equi}$ \\
\hline  & $f(\phi)\sim \xi\phi^2$ & $f(\phi)\sim \xi\phi^4$& $f(\phi)\sim \xi\phi^2$ & $f(\phi)\sim \xi\phi^4$ \\
\hline\\ $\phi$&
  $0.020<\xi<0.089$
& $0.017<\xi<0.089$&$0.018<\xi<0.107$ &$0.011<\xi<0.089$\\\\
$\phi^{2}$&   $0.015<\xi<0.084$
 & $0.012<\xi<0.090$ & $0.011<\xi<0.098$&$0.012<\xi<0.090$\\\\
$\phi^{3}$&   $0.018<\xi<0.085$
 & $0.008<\xi<0.083$ &$0.02<\xi<0.079$ &$0.019<\xi<0.080$\\\\
$\phi^{4}$&   $0.007<\xi<0.108$
 & $0.032<\xi<0.105$& $0.029<\xi<0.081$&$0.024<\xi<0.871$\\\\
$\phi^{\frac{2}{3}}$&  $0.04831<\xi$
 & $0.011<\xi<0.082$ &$0.086<\xi<0.096$&$0.007<\xi<0.095$\\\\
$e^{-\phi}$&   $0.013<\xi<0.088$
 & $0.043<\xi<0.078$&$0.025<\xi<0.077$ &$0.021<\xi<0.081$ \\\\
\hline\\\\
\end{tabular}
\end{center}
\end{tiny}
\end{table*}
\begin{figure*}
\flushleft\leftskip3em{
\includegraphics[width=.38\textwidth,origin=c,angle=0]{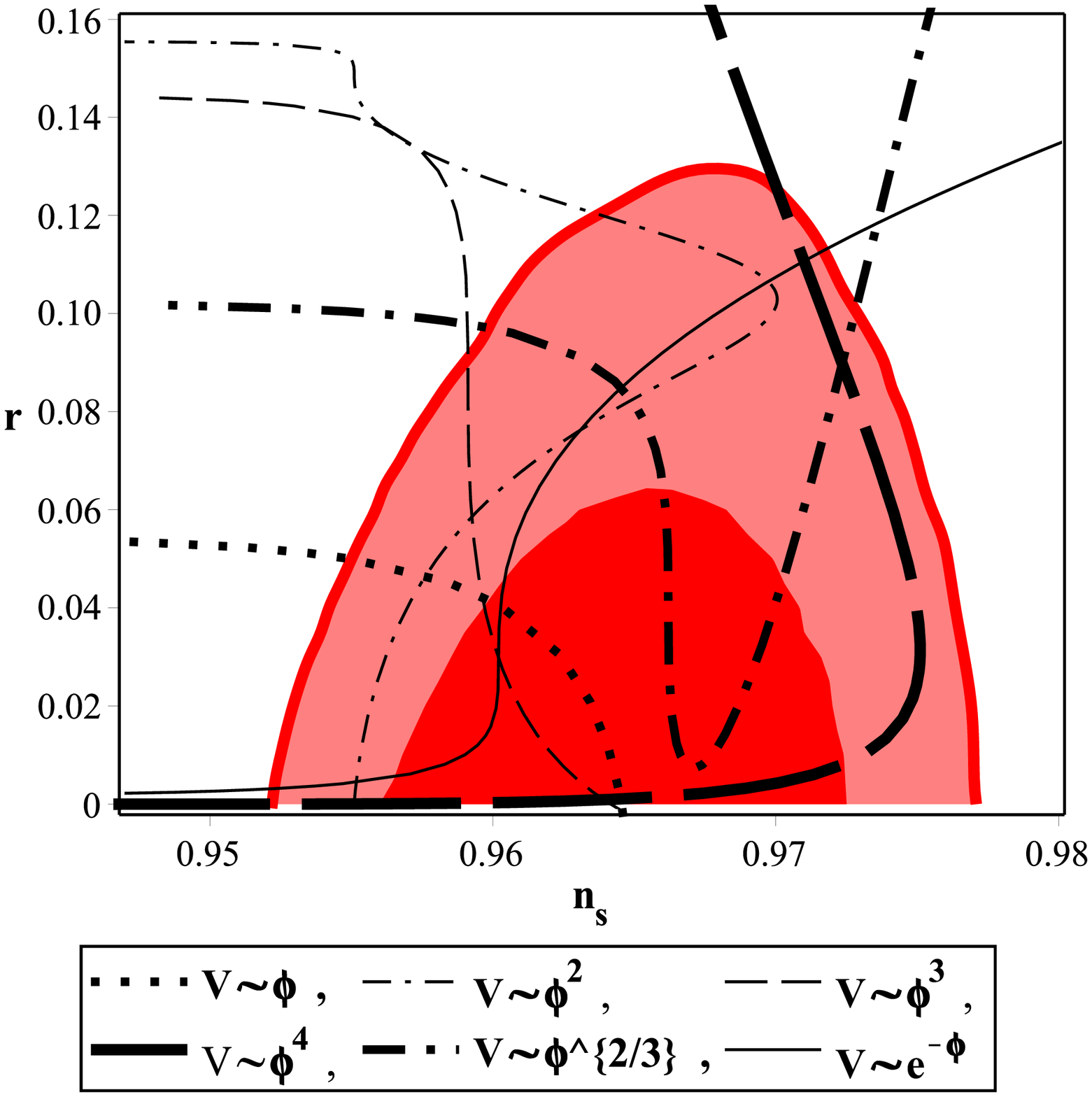}
\hspace{1cm}
\includegraphics[width=.38\textwidth,origin=c,angle=0]{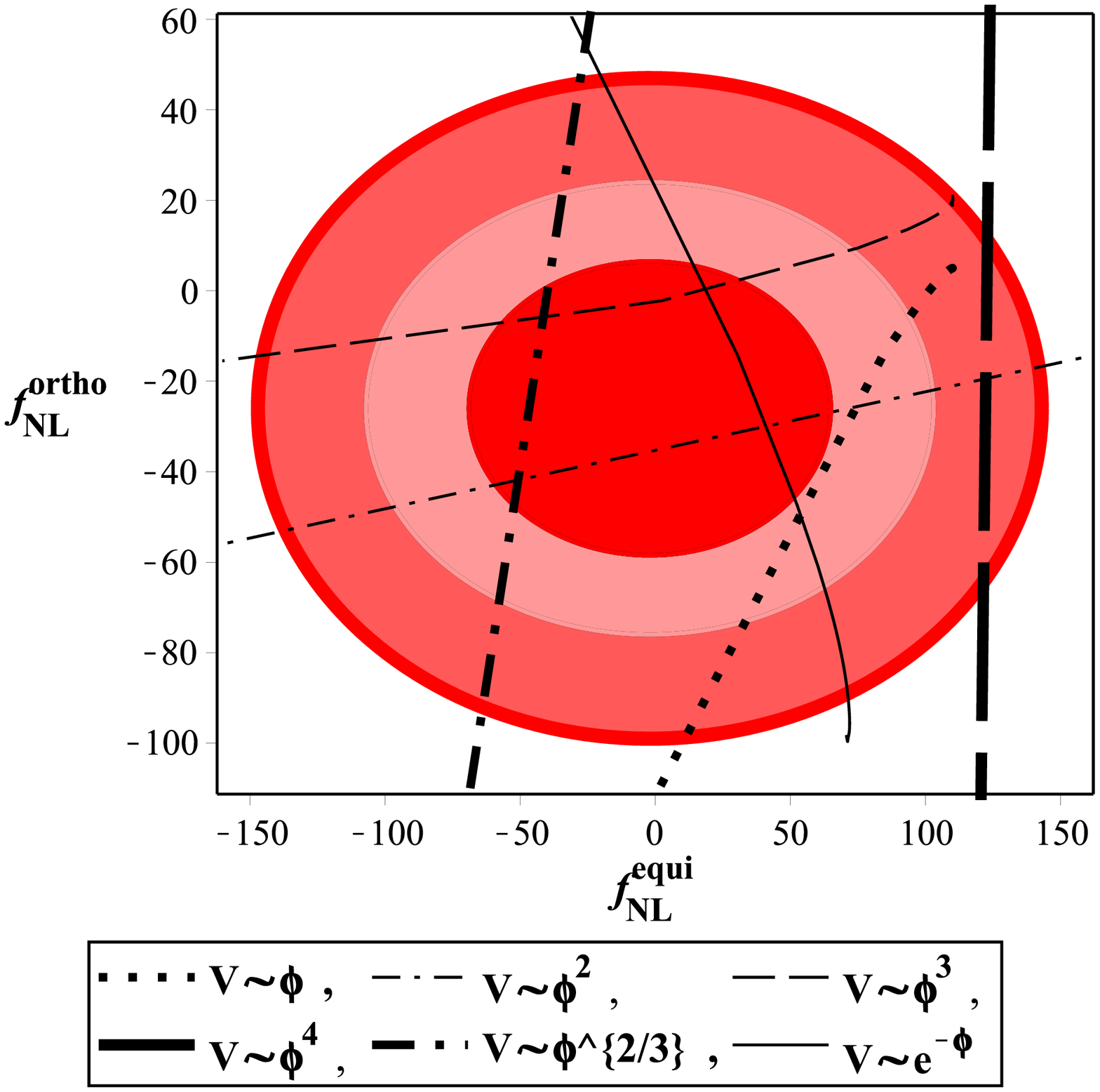}}
\caption{\label{fig13} Tensor-to-scalar ratio versus the scalar
spectral index for an inflationary model in which both inflaton and
its derivative are nonminimally coupled to gravity, in the
background of Planck2015 TT, TE, EE+lowP data (the left panel) and
the amplitude of the orthogonal configuration of non-Gaussianity
versus the amplitude of the equilateral configuration for an
inflationary model in which both inflaton and its derivative are
nonminimally coupled to gravity, in the background of Planck2015
TTT, EEE, TTE and EET data (right panel). Figure is plotted with
$n=4$, $f(\phi)\sim \xi\phi^2$ and various types of potential, for
 $N=60$.}
\end{figure*}
\begin{figure*}
\flushleft\leftskip3em{
\includegraphics[width=.38\textwidth,origin=c,angle=0]{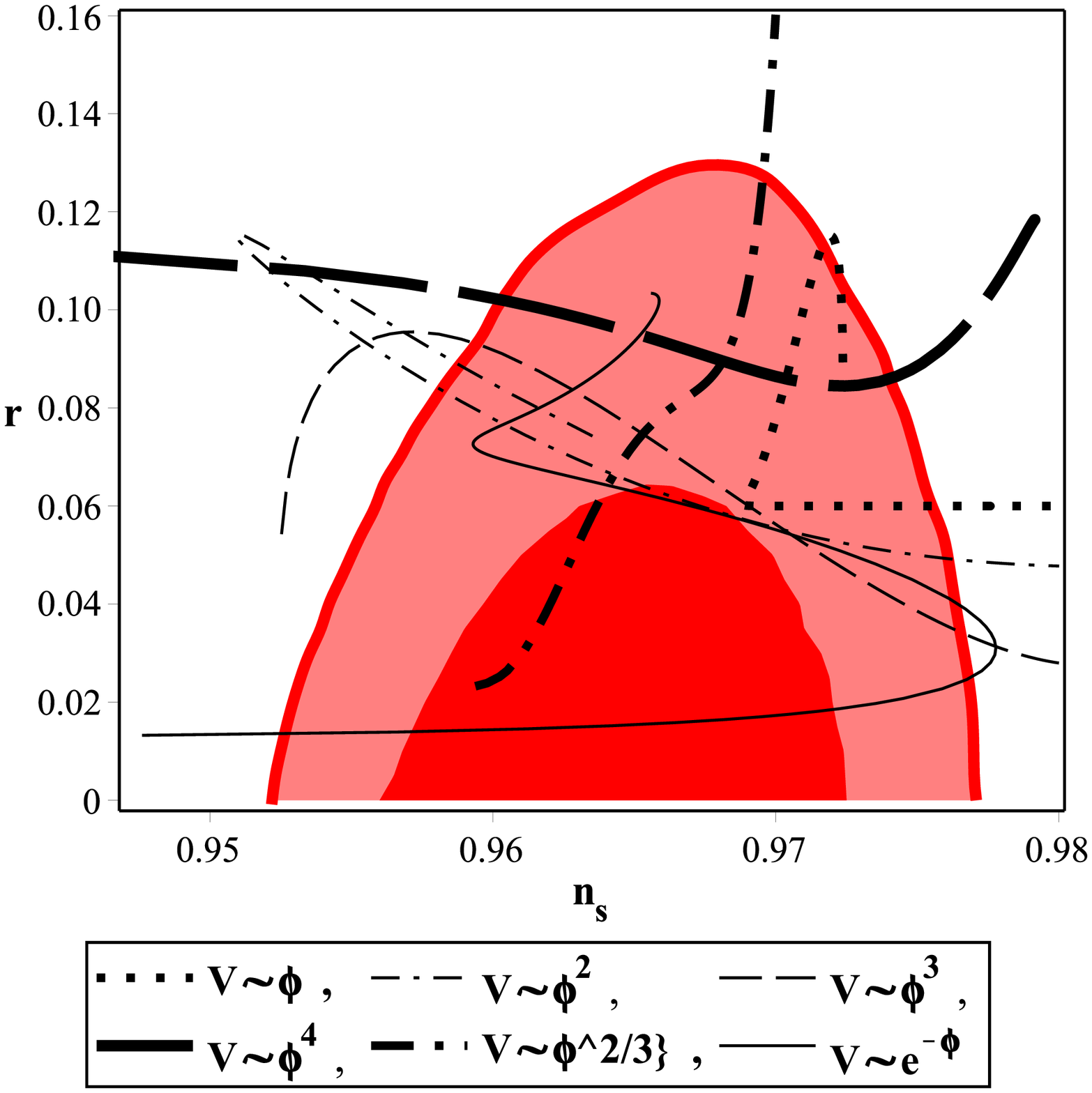}
\hspace{1cm}
\includegraphics[width=.38\textwidth,origin=c,angle=0]{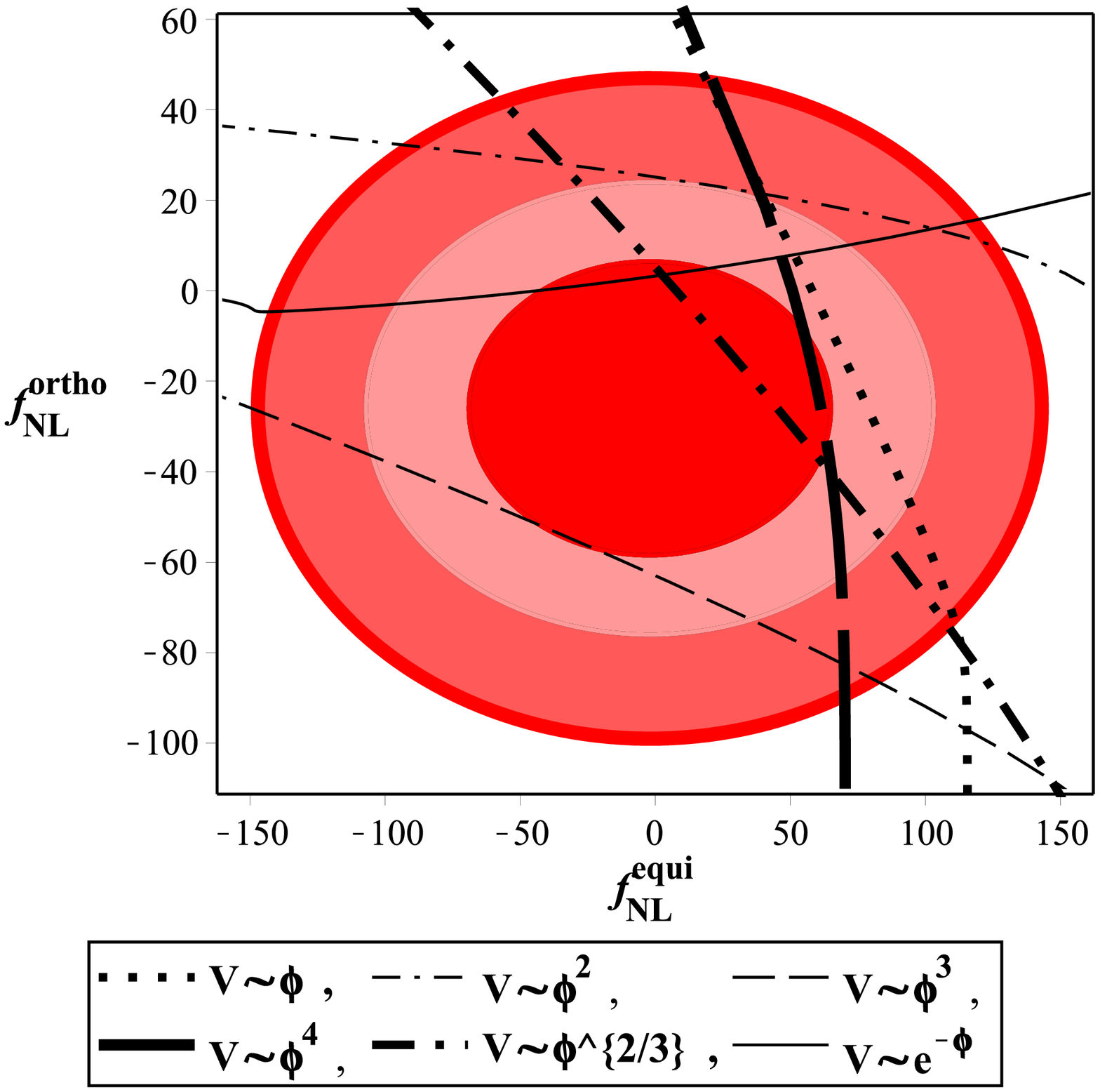}}
\caption{\label{fig15} Tensor-to-scalar ratio versus the scalar
spectral index for an inflationary model in which both inflaton and
its derivative are nonminimally coupled to the gravity, in the
background of Planck2015 TT, TE, EE+lowP data (left panel) and the
amplitude of the orthogonal configuration of non-Gaussianity versus
the amplitude of the equilateral configuration for an inflationary
model in which both inflaton and its derivative are nonminimally
coupled to the gravity, in the background of Planck2015 TTT, EEE,
TTE and EET data (right panel). Figure is plotted with $n=4$,
$f(\phi)\sim\xi\phi^4$ and various types of potential, for $N=60$.}
\end{figure*}
\begin{table*}
\begin{tiny}
\begin{center}
\caption{\label{tab:4} The ranges of $\xi$ in which the values of
the inflationary parameters $r$ and $n_{s}$ and also,
$f_{NL}^{ortho}$ and $f_{NL}^{equi}$, with $n=4$, are compatible
with the $95\%$ CL of the Planck2015 dataset.}
\begin{tabular}{ccccc}
\\ \hline \hline$V$& $r$ - $n_{s}$ & $r$ - $n_{s}$ &
 $f_{NL}^{ortho}$ - $f_{NL}^{equi}$ &  $f_{NL}^{ortho}$ - $f_{NL}^{equi}$ \\
\hline  & $f(\phi)\sim\xi\phi^2$ & $f(\phi)\sim\xi\phi^4$ & $f(\phi)\sim\xi\phi^2$ &$f(\phi)\sim\xi\phi^4$ \\
\hline\\ $\phi$&
 $\xi<0.084$
& $\xi<0.021$\,,\,$0.025<\xi<0.084$ &$\xi<0.113$ & $0.016<\xi<0.093$\\\\
$\phi^{2}$&  $\xi<0.083$
 & $\xi<0.040$\,,\,$0.059<\xi<0.112$ &$0.125<\xi<0.090$ &$0.012<\xi<0.092$\\\\
$\phi^{3}$&  $\xi<0.082$
 & $0.019<\xi<0.087$ &$\xi<0.106$ &$0.007<\xi<0.109$\\\\
$\phi^{4}$&   $0.075<\xi<0.105$
 & $0.019<\xi<0.081$&$0.036<\xi<0.079$ &$0.019<\xi<0.0.935$\\\\
$\phi^{\frac{2}{3}}$& $0.043<\xi<0.112$
 & $\xi<0.086$  &$0.009<\xi<0.099$& $0.008<\xi<0.098$\\\\
$e^{-\phi}$&  $0.012<\xi<0.099$
 & $\xi<0.109$&$0.004<\xi<0.104$ & $0.013<\xi<0.096$\\\\
\hline\\\\
\end{tabular}
\end{center}
\end{tiny}
\end{table*}

\section{Summary}
In this paper, we have studied the dynamics of a generalized
inflationary model in which both the scalar field and its
derivatives are nonminimally coupled to gravity. In order to have a
complete treatment of the model, we have considered an extension of
the nonminimal derivative coupling in the lagrangian as
${\cal{F}}(\phi)G_{\mu\nu}\nabla^{\mu}\nabla^{\nu}(\phi)$ where
${\cal{F}}$ is a general function of $\phi$. We have firstly
obtained the main equations of this generalized inflationary model.
Then, using the ADM formulation of the metric, we have studied the
linear perturbations of the model. We have expanded the action up to
the second order in perturbation and by using the 2-point
correlation function, we have derived the amplitude of the scalar
perturbation and its spectral index in our setup. To study the
tensor perturbation we have used the tensor part of the perturbed
metric and we have calculated the tensor perturbation and its
spectral index too. The ratio between the amplitude of the tensor
and scalar perturbations has been obtained in our setup. After that,
in order to seek the non-Gaussian feature of the primordial
perturbations, we have studied the non-linear theory. To study the
non-linear perturbation, one has to calculate the three point
correlation functions. In this regard, we have expanded the action
of the model up to the third order in perturbation. Then, by using
the interacting picture we have computed the three point correlation
function and the nonlinearity parameter in our model. Also, by
introducing the shape functions as $\breve{{\cal{S}}}^{equi}$ and
$\breve{{\cal{S}}}^{ortho}$, we have found the amplitude of the
non-Gaussianity in the equilateral and orthogonal configurations.
Since, both the equilateral and orthogonal configuration have peek
in the limit $k_{1}=k_{2}=k_{3}$, we have focused in this limit.
After finding the important perturbation parameters, we have
analyzed our model with recently released Planck 2015 observational
data. To this end, we had to specify general functions of the scalar
field in the model. For the the extension of the nonminimal
derivative term, we have adopted the function to be as
${\cal{F(\phi)}}\sim\frac{1}{2n}\phi^n$ with $n=1, 2, 3, 4$. We have
also defined two functions for the nonminimal coupling between the
scalar field and the curvature scalar as $f\sim \xi\phi^{2}$ and
$f\sim \xi\phi^{4}$. Then, by adopting several types of potential as
linear, quadratic, cubic, quartic potential, a potential
corresponding to $\phi^{\frac{2}{3}}$ and an exponential potential,
we have analyzed this inflationary models numerically and compared
our model with the observational data of Planck2015. In this regard,
we have studied the behavior of the tensor-to-scalar ratio versus
the scalar spectral index in the background of the Planck2015 TT,
TE, EE+lowP data and obtained some constraints on the nonminimal
coupling parameter. We have also analyzed the non-Gaussinaty feature
of the primordial perturbations numerically by studying the behavior
of the orthogonal configuration versus the equilateral configuration
in the background of the Planck2015 TTT, EEE, TTE and EET data. In
this paper, we have shown that this generalized inflationary model
is observationally viable in some ranges of its parameters space. As
an important result, we have shown that the nonminimal couplings
help to make models of chaotic inflation, that would otherwise be in
tension with Planck data, in better agreement with the data.

In summary, as an important result, our numerical analysis shows
that this non-minimal model, in some ranges of the non-minimal
coupling parameter and with small values of this parameter (the weak
coupling limit), is consistent with observation. Therefore, this
model in weak coupling limit has cosmological viability. Also, in
the weak coupling limit this setup is applicable to several
inflationary potentials. In this limit and in the ranges obtained by
numerical analysis, the scalar spectral index is red tilted. Another
point is that in our setup, in some range of the NMC parameter, it
is possible to have large non-Gaussianity.\\

\newpage
\bf{ Appendix: The expanded action up to third order} \label{A}

\begin{eqnarray}
S_{3}=\int dt\, d^{3}x\,
a^{3}\Bigg[\left(3\kappa^{-2}H^{2}(1+\kappa^{2}f)-\frac{\dot{\phi}^{2}}{2}+6H\dot{\phi}f'+30H^{2}\dot{\phi}^{2}{\cal{F}}'\right)\Phi^{3}
+\Phi^{2}\Bigg(\Big(\frac{3}{2}\dot{\phi}^{2}
-9\kappa^{-2}H^{2}(1+\nonumber\\
\kappa^{2}f) -18H\dot{\phi}f'
-54H^{2}\dot{\phi}^{2}{\cal{F}}'\Big){\Psi}+\Big(-6\kappa^{-2}H(1+\kappa^{2}f)
-6\dot{\phi}f'-36H\dot{\phi}^{2}{\cal{F}}'\Big)\dot{\Psi}
+2\dot{\phi}^{2}{\cal{F}}'\frac{\partial^{2}{\Psi}}{a^{2}}+\nonumber\\
\Big(2\kappa^{-2}H(1+\kappa^{2}f)
+12H\dot{\phi}^{2}{\cal{F}}'+2\dot{\phi}f'\Big)
\frac{\partial^{2}B}{a^{2}}\Bigg)+\Phi\Bigg(-a^{-2}\Big(2\kappa^{-2}H(1+\kappa^{2}f)\nonumber\\
+2\dot{\phi}f'
+6\dot{\phi}^{2}{\cal{F}}'\Big)\partial_{i}{\Psi}\partial_{i}B+9\Big(2\kappa^{-2}H(1+\kappa^{2}f)
+2\dot{\phi}f' +6\dot{\phi}^{2}{\cal{F}}'\Big)\dot{\Psi}
{\Psi}\nonumber\\-\frac{\kappa^{-2}(1+\kappa^{2}f)+3\dot{\phi}^{2}{\cal{F}}'}{2a^{4}}
\Big(\partial_{i}\partial_{j}B
\partial_{i}\partial_{j}B-\partial^{2}B
\partial^{2}B\Big)-
\frac{\Big(2\kappa^{-2}H(1+\kappa^{2}f) +2\dot{\phi}f'
+6\dot{\phi}^{2}{\cal{F}}'\Big)}{a^{2}}{\Psi}\partial^{2}B-
\nonumber\\
\frac{2\kappa^{-2}(1+\kappa^{2}f)+6\dot{\phi}^{2}{\cal{F}}'}
{a^{2}}\dot{\Psi}\partial^{2}B
-2\frac{\kappa^{-2}+f+\dot{\phi}^{2}{\cal{F}}'}{a^{2}}{\Psi}\partial^{2}{\Psi}
-\frac{\kappa^{-2}+f+\dot{\phi}^{2}{\cal{F}}'}{a^{2}}(\partial{\Psi})^{2}
\nonumber\\+\Big(3\kappa^{-2}(1+\kappa^{2}f)+9\dot{\phi}^{2}{\cal{F}}'
\Big)\dot{\Psi}^{2}\Bigg)+\frac{\kappa^{-2}+f-\dot{\phi}^{2}{\cal{F}}'}{a^{2}}{\Psi}(\partial{\Psi})^{2}
-9\Big(\kappa^{-2}+f+\dot{\phi}^{2}{\cal{F}}'\Big)\dot{\Psi}^{2}{\Psi}\nonumber\\
+2\frac{\kappa^{-2}+f+\dot{\phi}^{2}{\cal{F}}'}{a^{2}}\dot{\Psi}\partial_{i}
{\Psi}\partial_{i}B+2\frac{\kappa^{-2}+f+\dot{\phi}^{2}{\cal{F}}'}{a^{2}}\dot{\Psi}
{\Psi}\partial^{2}B
\nonumber\\
+\frac{3}{2}\frac{\Big(\kappa^{-2}+f+\dot{\phi}^{2}{\cal{F}}'\Big){\Psi}\Big(\partial_{i}\partial_{j}B
\partial_{i}\partial_{j}B-\partial^{2}B
\partial^{2}B\Big)}{a^{4}}-2\frac{\kappa^{-2}+f+\dot{\phi}^{2}{\cal{F}}'}{a^{4}}\partial_{i}{\Psi}\partial_{i}B \partial^{2}B\Bigg]\nonumber
\end{eqnarray}


\end{document}